\documentclass[useAMS,usenatbib]{mn2e}
\usepackage{graphicx}
\usepackage{longtable}
\usepackage{amssymb}

\title[The Hard X-Ray Luminosity Function of $3<z\lesssim 5$ AGN]{The Hard X-Ray Luminosity Function of High-Redshift ($3<z\lesssim 5$) Active Galactic Nuclei}
\author[F. Vito et al.]
{F. Vito$^{1,2}$\thanks{E-mail: fabio.vito@unibo.it},
R. Gilli$^{2}$,
C. Vignali$^{1,2}$,
A. Comastri$^{2}$, 
M. Brusa$^{1,2,3}$, 
N. Cappelluti$^{2}$,
\newauthor
K. Iwasawa$^{4}$
\\ \\
$^{1}$ Dipartimento di Fisica e Astronomia, Universit\`a degli Studi di Bologna, Viale Berti Pichat 6/2, 40127 Bologna, Italy \\
$^{2}$ INAF -- Osservatorio Astronomico di Bologna, Via Ranzani 1, 40127 Bologna, Italy\\
$^{3}$ Max-Planck-Institut f\"{u}r extraterrestrische Physik (MPE), Giessenbachstrasse 1, D-85748, Garching bei M\"{u}nchen, Germany\\
$^{4}$ ICREA and Institut de Ciències del Cosmos (ICC), Universitat de Barcelona (IEEC-UB), Mart\'{i} i Franqu\`{e}s 1, 08028, Barcelona, Spain\\
Accepted 2014 September 23.  Received 2014 September 10; in original form 2014 August 5
}

\begin{document}

\date{}

\graphicspath{{.}}

\pagerange{\pageref{firstpage}--\pageref{lastpage}} \pubyear{2014}

\maketitle

\label{firstpage}

\begin{abstract}
We present the hard-band ($2-10\,\rmn{keV}$) X-ray luminosity function (HXLF) of $0.5-2\,\rmn{keV}$ band selected AGN at high redshift. We have assembled a sample of 141 AGN at $3<z\lesssim5$ from X-ray surveys of different size and depth, in order to sample different regions in the $ L_X - z$ plane. The HXLF is fitted in the range $\rmn{logL_X\sim43-45}$ with standard analytical evolutionary models through a maximum likelihood procedure. The evolution of the HXLF is well described by a pure density evolution, with the AGN space density declining by a factor of $\sim10$ from $z=3$ to 5. 
A luminosity-dependent density evolution model which, normally, best represents the HXLF evolution at lower redshift, is also consistent with the data, but a larger sample of low-luminosity ($\rmn{logL_X}<44$), high-redshift AGN is necessary to constrain this model. We also estimated the intrinsic fraction of AGN 
obscured by a column density $\rmn{logN_H}\geq23$ to be $0.54\pm0.05$, with no strong dependence on luminosity. This fraction is higher than the value in the Local Universe, suggesting an evolution of the luminous ($\mathrm{L_X>10^{44}\mathrm{erg\,s^{-1}}}$) obscured AGN fraction from $z=0$ to $z>3$.
\end{abstract}

\begin{keywords}
methods: data analysis -- surveys -- galaxies: active -- X-rays: galaxies
\end{keywords}

\section{Introduction}\label{sec1}

 The presence of Super Massive Black Holes (SMBH) with masses of the order of $\gtrsim10^9 \rmn{M_\odot}$ at $z>6$ \citep{Willott03, Fan06, Fan06b,Willott09,Mortlock11}, when the Universe was less than 1 Gyr old, challenges our knowledge of SMBH formation and growth. In order to reach such masses in a few hundreds million years, continuous Eddington-limited, or even super-Eddington, accretion is required by the two most promising models of SMBH seeds formation, i.e. remnant of POP III stars \citep[][see also \citealt{Whalen11}]{Madau01,Alvarez09,Volonteri10} and direct collapse of massive gas clouds \citep{Begelman10,Ball11}. The radiation produced by accretion process, traced by the presence of Active Galactic Nuclei (AGN), at this stage is expected to be severely obscured by the large amount of infalling gas, necessary to the build-up of the BH mass. 

Assessing the cosmic evolution of the AGN population up to very high redshifts is a key factor to understand and constrain the early growth of SMBH, but it requires the collection of large, complete and reliable samples of AGN. X-ray surveys, such as the 4 Ms \textit{Chandra} Deep Field South \citep[4 Ms CDFS,][]{Xue11}, XMM-COSMOS \citep{Hasinger07}, \textit{Chandra}-COSMOS \citep{Elvis09} and Subaru-XMM Deep Survey \citep[SXDS;][]{Ueda08} are the best tool to gather such samples, since they are less biased against obscuration and galaxy dilution than optical/NIR surveys \cite[e.g.][]{Brandt05}. Moreover, all these surveys, exploiting deep multiwavelength coverage, are characterised by a very-high ($>90$ per cent) redshift completeness.

Several works were performed on the evolution of the X-ray selected AGN population at $0\leq z\lesssim5$ \citep[i.e.][]{Miyaji00,Ueda03, Hasinger05,LaFranca05, Aird08, Silverman08, Ebrero09, Yencho09, Ueda14}. According to the majority of previous investigations, the evolution of the AGN space density is well described by a Luminosity-Dependent Density Evolution (LDDE). The AGN population experienced a ``downsizing" evolution in which the space density of luminous AGN ($L_X\gtrsim10^{45}\rmn{erg\,s^{-1}}$) peaked at higher redshift ($z\sim2-3$) than that of less luminous ones ($z\sim 1- 1.5$). This behaviour is similar to the ``cosmic downsizing" of galaxies \citep[e.g.][]{Cowie96, Thomas05, Damen09}. \cite{Aird10} proposed a Luminosity And Density Evolution (LADE) model, in which the  break luminosity and normalization evolve independently with redshift. However, all of these works preferentially constrain the evolution at $z\lesssim3$, because of the lack of sizeble samples of high-redshift, X-ray selected 
AGN with a high completeness level. 

A few authors focused on the X-ray selected $z>3$ AGN population. In particular, \cite{Brusa09} and \cite{Civano11}, using data from the XMM-Newton and \textit{Chandra} surveys, respectively, in the COSMOS field, presented the space density of high-redshift AGN. \cite{Hiroi12} extended these works by adding data from the SXDS field. All of their results are consistent with a decline in the AGN space density up to $z\sim 5$ similar to the evolution of optically selected quasars \citep[QSOs, e.g.][]{Schmidt95}. The same conclusions were reached by analysing the AGN number counts (\citealt{Brusa09}, \citealt{Civano11}, \citealt{Lehmer12} and \citealt{Vito13} in the 4 Ms CDFS). 
However, single X-ray surveys provide only relatively small samples of high-redshift AGN (tens of objects, see e.g. \citealt{Trichas12}), making the evolution of the $z>3$ AGN population poorly constrained. For instance, in contrast to the above mentioned works, \cite{Fiore12} reported that a pure luminosity evolution (PLE) is the best model for describing the luminosity function of their X-ray and optically selected sample of $z>3$ AGN in the 4 Ms CDFS. Recently, \cite{Kalfountzou14} presented the number counts and space density of the largest sample (209 objects) of X-ray selected AGN at $z\gtrsim3$, using data from the \textit{Chandra}-COSMOS and ChaMP \citep{Kim07,Green09} surveys. Their findings are consistent with the declining space density scenario.

Besides the statistics, one important source of uncertainties in assessing the evolution of the AGN population is obscuration. 
X-ray background synthesis models \citep{Gilli07, Treister09, Akylas12} predict that the majority of AGN is obscured by a column density of $N_H>10^{22}\rmn{cm^{-2}}$. 
Currently, the fraction of obscured, especially Compton Thick ($N_H>10^{24}\rmn{cm^{-2}}$) AGN is well constrained only in the Local Universe \citep[e.g.][]{Burlon11}. An anti-correlation between the obscured AGN fraction and the luminosity seems to be in place up to $z\sim3$ \citep[e.g.][]{Steffen03,Ueda03,LaFranca05,Treister06,Hasinger08, Ebrero09, Ueda14}. Moreover, evidences were found for a positive evolution with redshift up to $z\sim2$ \citep[e.g.][; but see also \citealt{Ueda03, Gilli10}]{LaFranca05, Akylas06, Treister06, Ueda14}, after which the trend seems to saturate \citep{Hasinger08}. Other evidences of a larger fraction of obscured AGN at $z\gtrsim2$ compared with the Local Universe were recently presented by \cite{Iwasawa12} and \cite{Hiroi12}. This behaviour can be interpreted by invoking two modes of accretion are responsible for the triggering of nuclear activity \citep[e.g.][]{Hasinger08,Hopkins08,Hickox09}. On the one hand, according to this scenario, luminous ($L_X\gtrsim10^{44}\rmn{erg\,
s^{-1}}$) AGN are 
triggered by wet (i.e. gas rich) mergers \citep[e.g.][]{DiMatteo05,Menci08} and nuclear activity occurs in short burst ($\sim0.01 \rmn{Gyr}$; e.g. \citealt{Alexander05}) of chaotic accretion \citep{Hopkins08}, which may produce large covering factors. The powerful radiation emitted then rapidly sweeps away the obscuring material and the QSO reveals itself as unobscured. Since the merger rate and, most importantly, the gas fraction increase with redshift \citep[and reference therein]{Carilli13}, a longer obscured phase and/or a larger covering factor is expected  at higher redshift, causing a positive evolution of the fraction of obscured AGN. On the other hand, low luminosity AGN are thought to be preferentially triggered by secular processes \citep[e.g.][]{Elbaz11}. Accretion in this regime is thought to be ``symmetric'' and obscuration would be only due to the geometry of the system with respect to the line of sight, as postulated by the \cite{Urry95} unification scheme. No strong redshift dependency is 
then expected for the fraction of obscured AGN in this luminosity regime. Constraints at 
high redshift on the evolution of obscuration are fundamental to test this framework.
 
In this work, we present a sample of 141 $3<z\lesssim5$ soft-band ($0.5-2\,\rmn{keV}$) X-ray selected AGN. The sample is collected from the 4 Ms CDFS, \textit{Chandra}-COSMOS, XMM-COSMOS and SXDS surveys and is characterised by a very-high ($\sim98\%$) redshift completeness (see the following section and Tab.~\ref{tab0}). We derived the luminosity function and fitted it with widely adopted evolutionary models in literature through a Maximum Likelihood procedure on unbinned data. Then, we estimated the fraction of obscured AGN and assessed the evolution of the AGN space density. Throughout this work we assume a $\Omega_m = 0.27$, $\Omega_\Lambda = 0.73$ and $H_0 = 70\, \rmn{km}\,\rmn{ s^{-1} Mpc^{-1}}$ Cosmology.

\section[]{The high-redshift sample}\label{sec2}
We collected a sample of 141 AGN detected in the soft X-ray band ($0.5-2\,\rmn{keV}$) in the redshift range $3<z<5.1$.
The $0.5-2\,\rmn{keV}$ band approximatively correspond to the $2-10\,\rmn{keV}$ band at $z\approx 3$. 
The sample is assembled from four surveys  of different sizes and depths. This is crucial to cover as homogeneously as possible the $ \rmn{L_X} -z$ plane, since distant, Seyfert-like AGN can be detected only by the deepest pencil-beam surveys, while the rare, very luminous QSOs need wide (and hence shallower) surveys to be found.
The choice of the fields was driven by the presence of massive multi-wavelength campaigns, which resulted in a very-high ($\gtrsim95 \%$) redshift completeness of the parent samples. We used both spectroscopic (78 objects, $\sim55$ per cent of the entire sample) and photometric (63 objects, $\sim45$ per cent) redshifts. Typical errors on the photometric redshifts for high-redshift sources without spectroscopic redshift are of the order of $\Delta z_{phot}/(1+z_{phot})\sim0.02$, therefore we do not expect the uncertainties on the photometric redshifts to strongly affect the results presented in the following sections. In \S~\ref{method} we will assign the sources to two different, absorption-based subsamples, where the threshold is set to $\rmn{logN_H}=23$. In the following sub-sections, we describe each of the considered surveys as well as the sample of high-redshift AGN collected from them.
We also quote the number of sources with no redshift information, which will be used in \S~\ref{complete} and the number of sources obscured by a column density $\rmn{logN_H}>23$ (see also Tab.~\ref{tab0}).

\subsection{The 4 Ms \textit{Chandra} Deep Field South (CDF-S)}\label{cdfs_sample}
The 4 Ms CDF-S \citep{Xue11} is the deepest X-ray survey to date, covering an area of $464.5\, \rmn{arcmin^2}$ with flux limit of 
$F\approx9.1\times10^{-18}\,\rmn{erg\,cm^{-2}s^{-1}}$ in the $0.5-2\,\rmn{keV}$ band.

In this work, we use the sample of $z>3$ X-ray selected AGN described by \cite{Vito13}, restricted to the redshift range $3<z<5.1$ by discarding 3 objects which are associated to photometric redshifts $5.1<z\la7.6$. Accounting only for the soft-band detected objects, the sample consists of 27 sources with intrinsic (i.e. rest-frame and corrected for absorption) luminosities $8\times10^{42}\la \frac{L_{2-10\,\rmn{keV}}}{\rmn{erg\,s^{-1}}}\la 10^{45}$. A spectroscopic redshift is available for 13 objects ($\sim48$ per cent), while a photometric redshift is associated to the remaining 14 sources. As for the redshift completeness, \cite{Vito13} showed that 39 out of the 740 ($\sim5.3$ per cent) X-ray sources in the survey main catalogue \citep{Xue11} have neither a spectroscopic nor a photometric redshift. This number decreases to $\sim4.8$ per cent (31 out of 650), if only the sources detected in the soft band are considered.

A spectral analysis was performed on this sample \citep{Vito13}, assuming an absorbed power-law with photon index fixed to $\Gamma=1.8$ and accounting for Galactic absorption ($N_H=7\times10^{19}\rmn{cm^{-2}}$). More complicated models could not be used because of the poor photon counting statistics characterising the sample. Sixteen out of 27 sources ($\sim63$ per cent) are highly obscured (best-fitting column density $N_H>10^{23}\rmn{cm^{-2}}$). The spectral parameters are reported in Tab.~\ref{tab1} and, more extensively, in \cite{Vito13}.

\subsection{The Subaru/XMM-\textit{Newton} Deep Survey (SXDS)}\label{SXDS}

The SXDS field (\citealt{Furusawa08,Ueda08, Hiroi12}) was observed with XMM-\textit{Newton} over an area of $1.14\,\rmn{deg^2}$
at a $0.5-2\,\rmn{keV}$ band flux limit of $F\approx6\times10^{-16}\,\rmn{erg\,cm^{-2}s^{-1}}$.

 The AGN candidates detected in the soft band are 733, and $\sim99.2$ per cent of them is associated to a spectroscopic or photometric redshift. Eighty-five X-ray sources are not covered by optical/near-IR observation and are not included in these numbers \citep{Hiroi12}.

\cite{Hiroi12} selected a sample of 30 AGN detected at $z>3$ in the soft band, making use of the spectroscopic (20 objects) and photometric (10 objects) information available in that field, and studied the comoving space density and the fraction of obscured AGN.

In their analysis, the intrinsic spectrum is assumed to be a power-law with photon index fixed to $\Gamma=1.9$ plus a contribution from a reflection component. We note that this shape mimics a spectrum with an effective photon index of $\Gamma\approx1.8$.
The sample lies in the redshift range $3<z\le4.5$ and is characterised by intrinsic luminosities in the range $10^{44}\la \frac{L_{2-10\,\rmn{keV}}}{\rmn{erg\,s^{-1}}}\la 10^{45}$. For each source, they derived an estimate of the column density by mean of the hardness ratio (HR):

\begin{equation}
 \rmn{HR} = \frac{\rmn{H-S}}{\rmn{H+S}}
\end{equation}
where S and H are the count rates in the $0.3-1.0$ and $1.0-4.5\,\rmn{keV}$ band, respectively. 
One source is obscured by a column density $N_H>10^{23}\rmn{cm^{-2}}$. In this work, we use the \cite{Hiroi12} sample and spectral parameters (shown in Tab.~\ref{tab1}).

\subsection{The \textit{Chandra} Cosmological Evolution Survey (C-COSMOS)}\label{C-COSMOS}
The C-COSMOS (\citealt{Elvis09,Puccetti09,Civano12}) covers an area of $\sim0.92\,\rmn{deg^2}$ with 
a total exposure time of $1.8$ Ms and a fairly uniform depth. The flux limit is
$F\approx1.9\times10^{-16}\,\rmn{erg\,cm^{-2}s^{-1}}$ in the $0.5-2\,\rmn{keV}$ band.
The number of detected X-ray sources is 1761.

Making use of the new redshifts available in this field \citep[][Lilly et al. in prep]{Salvato11, Civano12}, we refined the \textit{Chandra}-COSMOS high-redshift sample presented by \cite{Civano11} and found 62 soft-band detected objects at $3<z<5.1$.

Almost half of our sample (30 out of 62 AGN) was also detected by the XMM-\textit{Newton} survey in the COSMOS field, which encompasses the \textit{Chandra} observations. Since in the outskirts of the \textit{Chandra} field the sensitivity drops below the sensitivity of XMM-COSMOS, there are a number of sources detected by XMM-\textit{Newton} whose flux is below the \textit{Chandra} limit at that point. In order 1) to count each source only once and 2) not to lose the AGN detected only in XMM-COSMOS, we did not consider all the C-COSMOS field, but only the central region. The proper way to define it would be the region where \textit{Chandra} is more sensitive than XMM-\textit{Newton}. However, for simplicity during the evaluation of the sky-coverage (see \S~\ref{skycov}), we defined the new C-COSMOS area by looking at the overlapping XMM and \textit{Chandra} images and cutting off an external frame of the \textit{Chandra} field as narrow as possible, with an area of $\sim0.20\,\rmn{deg^2}$, to fulfil the two 
requirements mentioned above.

All the sources detected either only by \textit{Chandra}  or by both \textit{Chandra} and XMM-\textit{Newton} in the resulting $0.72\,\rmn{deg^2}$ region will then be counted in the C-COSMOS sample. Three objects (\textit{Chandra} ID 521, 1730 and 325) detected by both \textit{Chandra} and XMM-\textit{Newton}, fall outside the newly defined boundaries of C-COSMOS; therefore, they are counted in the XMM-COSMOS sample (XMM ID 5120, 5259 and 5347, respectively).  
Hence, the final high-redshift C-COSMOS sample consists of 59 AGN in the redshift range $3<z\le5.1$. A spectroscopic redshift is associated to 30 of them. We performed a spectral analysis assuming the same spectral model described in \S~\ref{cdfs_sample} (i.e. an absorbed power-law with $\Gamma=1.8$ and Galactic column density of $2.5\times10^{20}\rmn{cm^{-2}}$; \citealt{Kalberla05}) and fitted the spectra of the 59 sources using the Cash statistics \citep{Cash79} to estimate the best-fitting parameters (Tab.~\ref{tab1}). We derived intrinsic luminosities of $10^{43}\la \frac{L_{2-10\,\rmn{keV}}}{\rmn{erg\,s^{-1}}}\la 2\times10^{45}$ and a fraction of $\sim 32$ per cent of very obscured sources ($N_H>10^{23}\rmn{cm^{-2}}$). The poor spectral quality of \textit{Chandra} ID 3449 prevented us from investigating the effect of absorption and a simple power-law was assumed as spectral model in order to derive the intrinsic hard band luminosity. This source will be conservatively considered 
unobscured in the following sections.

 As for the redshift completeness, neither a spectroscopic nor a photometric redshift is available for 28 out of 1223 soft-band detected objects which fall on the reduced C-COSMOS area ($\sim 2$ per cent).

\subsection{The XMM-\textit{Newton} Cosmological Evolution Survey (XMM-Cosmos)}\label{XMM-COSMOS}
XMM-COSMOS (\citealt{Hasinger07,  Brusa10, Cappelluti09}) covers an area of $ 2.13\,\rm{deg^2}$ for a total of $\sim 1.55\, \rm{Ms}$. 
We considered as parent sample the 1797 sources reported in the \cite{Brusa10} multi-wavelength catalogue. Redshift were collected from \cite{Brusa10}, \cite{Salvato11} and \cite{Civano12} 
 and 53 soft-band detected objects resulted to be at $z>3$. 
 
 \cite{Brusa09} presented results on a sample of high-redshift AGN in XMM-COSMOS. In particular, they selected 39 soft-band detected AGN $z>3$ with $F_{0.5-2\,\rmn{keV}}>10^{-15}\,\rmn{erg\,cm^{-2}s^{-1}}$. Because of the new redshift information we collected, these two samples have 34 sources in common.

Once the central reduced C-COSMOS field is cut out, the remaining XMM-COSMOS area is $\sim 1.45\,\rm{deg^2}$ and the number of soft-detected object that are placed in that frame are 930, with a redshift completeness of $99.4$ per cent (i.e. 6 objects have no redshift information). Twenty-five objects are high-redshift sources (19 of which are included in the \cite{Brusa09} sample); 15 of them have a spectroscopic redshift. 
 All but one of the remaining 28 high-redshift AGN, which fall on the reduced C-COSMOS area, are detected by Chandra and are counted in the C-COSMOS sample (see \S~\ref{C-COSMOS})\footnote{XID 54039 is a high-redshift X-ray source which falls on the central, reduced Chandra field, but is detected only by XMM. This is probably due to statistical fluctuations affecting the flux of the source or the nearby background, or even to intrinsic variability of the source. Since in the reduced C-COSMOS region we considered only sources detected by \textit{Chandra}, we excluded it from the high-redshift sample.}.
 
\cite{Mainieri07,Mainieri11} performed a detailed spectral analysis on a subsample of X-ray sources in XMM-COSMOS, including 16 objects in our high-redshift sample. They assumed an absorbed power-law with photon index free to vary. Only one of them resulted to be heavily obscured (XID 2518, $\rmn{logN_H}=23.07^{+0.35}_{-0.37}$), with a very steep photon index ($\Gamma>2$). Since the photon counting statistics was too poor to perform a spectral analysis on all these objects and in order to be consistent with the spectral assumptions used for the other surveys and with the procedure adopted in \S~\ref{method}, we conservatively consider all the sources in our $z>3$ sample as unabsorbed AGN. This is also justified since, at similar fluxes, the SXDS sample (see \S~\ref{SXDS}) included only unabsorbed object, with just one exception. We derived the rest-frame hard-band luminosities from the soft-band flux reported in \cite{Cappelluti09}, assuming a simple power-law with $\Gamma=1.8$ (Tab.~\ref{tab1}).

\begin{table}
\caption{Main properties of the individual surveys}\label{tab0}
\begin{tabular}{|r|r|r|r|r|r|r|r|}
\hline
  \multicolumn{1}{|r|}{ FIELD} &
  \multicolumn{1}{|r|}{{\bf $\rmn{A\,[deg^2] }$ }} &
    \multicolumn{1}{r|}{{\bf $\%_{\rmn{compl}}$}} &
  \multicolumn{1}{r|}{{\bf $\rmn{N\,(N_{abs})}$}} &
  \multicolumn{1}{r|}{{\bf $\%_{\rmn{zspec}}$}} \\
      (1)   &  (2)   & (3)  & (4)  &  (5)    \\

  \hline
CDF-S            & 0.129 & 95 &  27 (16) & 48 \\
C-COSMOS$_{red}$ & 0.717 & 98 &  59 (19) & 51 \\
XMM-COSMOS$_{red}$ & 1.453 & 99 &  25 (0)  & 60 \\
SXDS             & 1.006 & 99 &  30 (1)  & 67 \\
 TOTAL           & 3.305 & 98 & 141 (36) & 55 \\
  \hline

\end{tabular}
(1) Field; (2) nominal area; (3) redshift completeness of the soft-band detected parent sample; (4) number of soft-band detected AGN at $3<z<5.1$, the number of sources included in the A subsample ($\rmn{logN_H}>23$; see \S~\ref{skycov}) is between brackets; (5) fraction of high-redshift AGN with a spectroscopic redshift. These numbers refer to the reduced field for \textit{Chandra} and XMM-COSMOS  and to the overlapping region between the X-ray and optical surveys for the SXDS field (see \S~\ref{sec2}.)

\end{table}

\begin{table*} 
\caption{Main information and spectral parameters of high-redshift sample.}
\label{tab1}
\begin{tabular}{|r|r|r|r|r|r|r|r|r|r|}
\hline
  \multicolumn{1}{c|}{ID} &
  \multicolumn{1}{c|}{RA} &
  \multicolumn{1}{c|}{DEC} &
  \multicolumn{1}{c|}{zadopt} &
  \multicolumn{1}{c|}{ztype} &
  \multicolumn{1}{c|}{zref} &
  \multicolumn{1}{c|}{$N_H$} &
  \multicolumn{1}{c|}{$F_{0.5-2\,\rmn{keV}} $} &
  \multicolumn{1}{c|}{$L_{2-10\,\rmn{keV}} $} \\

  \multicolumn{1}{c}{(1)} &
  \multicolumn{1}{c}{(2)} &
  \multicolumn{1}{c}{(3)} &
  \multicolumn{1}{c}{(4)} &
  \multicolumn{1}{c}{(5)} &
  \multicolumn{1}{c}{(6)} &
  \multicolumn{1}{c}{(7)} &
  \multicolumn{1}{c}{(8)} &
  \multicolumn{1}{c}{(9)}\\
\hline
  \multicolumn{9}{c}{C-COSMOS}\\
  \hline
43    & 150.18087 & 2.075997 & 3.01  & 1 & 1  & $<3$                 & 1.41 & 1.49\\
64    & 150.36472 & 2.143831 & 3.328 & 1 & 1  & $<7$                 & 3.12 & 4.80\\
75    & 150.24770 & 2.442225 & 3.029 & 1 & 1  & $22_{-7}^{+9}$       & 2.57 & 8.35\\
83    & 150.21418 & 2.475118 & 3.075 & 2 &  1 & $17_{-7}^{+9}$       & 2.00 & 5.52\\
113   & 150.20884 & 2.482010 & 3.333 & 1 & 1  & $<4$                 & 3.04 & 4.03\\
  \smallskip
124   & 150.20536 & 2.502848 & 3.072 & 2 &  1 & $13_{-10}^{+15}$     & 1.71 & 4.14\\
270   & 150.10734 & 1.759256 & 4.16  & 1 & 1  & $<72$                & 0.70 & 1.84\\
308   & 149.73615 & 2.179954 & 4.255 & 1 & 1  & $<30$                & 1.24 & 4.05\\
349   & 150.00438 & 2.038978 & 3.515 & 1 & 1  & $<5$                 & 1.62 & 2.43\\
386   & 150.37885 & 1.876099 & 3.33  & 2 &  1 & $21_{-19}^{+33}$     & 0.63 & 2.19\\
407   & 149.80849 & 2.313858 & 3.471 & 2 &  1 & $<4$                 & 2.67 & 3.88\\
  \smallskip

413   & 149.86968 & 2.294064 & 3.345 & 1 & 1  & $21_{-8}^{+10}$      & 2.71 & 9.05\\
472   & 149.96920 & 2.304833 & 3.155 & 1 & 1  & $17_{-10}^{+16}$     & 1.24 & 3.57\\
507   & 149.85845 & 2.409299 & 4.108 & 2 & 1  & $15_{-7}^{+8}$       & 4.89 & 18.50\\
529   & 149.98156 & 2.315056 & 3.017 & 1 & 2  & $17_{-7}^{+9}$       & 1.93 & 5.13\\
558   & 149.88247 & 2.505174 & 3.1   & 2 & 1  & $4_{-3}^{+4}$        & 4.34 & 6.90\\
688   & 150.34505 & 1.958014 & 3.065 & 2 & 1  & $<23$                & 0.84 & 1.66\\
      \smallskip
689   & 150.41521 & 1.934286 & 3.681 & 2 & 1  & $<16$                & 0.86 & 1.43\\
691   & 149.81217 & 2.282920 & 3.297 & 2 & 1  & $<6$                 & 0.90 & 1.16\\
693   & 149.85148 & 2.276539 & 3.371 & 1 & 1  & $<7$                 & 1.35 & 1.85\\
700   & 149.85151 & 2.426858 & 3.35  & 2 &  1 & $46_{-23}^{+33}$     & 5.43 & 3.52\\
781   & 150.10093 & 2.419495 & 4.66  & 1 & 1  & $<61$                & 0.74 & 4.14\\
784   & 150.30071 & 2.300688 & 3.498 & 1 & 1  & $<76$                & 0.33 & 1.21\\
      \smallskip
815   & 150.00937 & 1.852672 & 4.032 & 2 &  1 & $<21$                & 0.71 & 1.45\\
879   & 150.38347 & 2.074682 & 3.859 & 1 & 2 & $76_{-70}^{+132}$    & 0.16 & 1.96\\
890   & 149.91958 & 2.345473 & 3.021 & 1 & 1 & $<14$                & 1.06 & 1.19\\
917   & 150.19256 & 2.219909 & 3.09  & 1 & 1 & $<20$                & 0.87 & 1.76\\
931   & 150.35964 & 2.073574 & 4.917 & 1 & 1 & $<43$                & 0.62 & 2.39\\
947   & 150.29719 & 2.148829 & 3.328 & 1 & 1 & $16_{-12}^{+29}$     & 0.56 & 1.65\\
      \smallskip
953   & 150.21070 & 2.391473 & 3.095 & 1 & 1 & $<22$                & 0.90 & 1.37\\
955   & 150.20899 & 2.438581 & 3.715 & 1 & 1 & $<6$                 & 1.05 & 1.78\\
965   & 150.15215 & 2.307818 & 3.175 & 1 & 1 & $104_{-42}^{+50}$    & 0.26 & 5.39\\
1040  & 150.22593 & 1.799779 & 3.264 & 2 & 1 & $52_{-27}^{+37}$     & 0.49 & 3.63\\
1118  & 149.87920 & 2.225811 & 3.65  & 1 & 1 & $<13$                & 1.72 & 3.68\\
1197  & 149.89429 & 2.433144 & 3.382 & 1 & 1 & $<12$                & 0.27 & 0.37\\
    \smallskip
1236  & 149.84572 & 2.481628 & 3.375 & 2 & 1 & $<28$                & 0.74 & 1.20\\
1263  & 150.42519 & 2.312089 & 3.092 & 2 & 1 & $59_{-40}^{+70}$     & 0.24 & 2.03\\
1269  & 150.54647 & 2.224128 & 3.506 & 2 & 1 & $12_{-9}^{+12}$      & 2.25 & 6.07\\
1303  & 149.99044 & 2.297347 & 3.026 & 1 & 1 & $<24$                & 0.82 & 1.43\\
1311  & 150.25977 & 2.376108 & 3.717 & 1 & 1 & $<35$                & 0.67 & 1.37\\
1392  & 150.45489 & 1.967361 & 3.485 & 1 & 1 & $<47$                & 0.72 & 1.73\\
    \smallskip
1490  & 149.80460 & 2.118866 & 3.791 & 2 & 1 & $<52$                & 0.34 & 1.02\\
1505  & 150.09688 & 2.021498 & 3.546 & 1 & 1 & $<20$                & 0.69 & 1.05\\
1509  & 150.31788 & 2.004926 & 3.428 & 2 & 1 & $<40$                & 0.58 & 1.21\\
1514  & 150.08617 & 2.138865 & 5.045 & 2 & 1 & $94_{-55}^{+73}$     & 0.30 & 5.14\\
1654  & 150.26739 & 1.700929 & 3.412 & 2 & 1 & $<11$                & 0.96 & 1.25\\
1656  & 150.27158 & 1.613616 & 3.466 & 2 & 1 & $32_{-28}^{+35}$     & 1.26 & 6.10\\

\hline\end{tabular}

 (1) source identification number as in \cite{Elvis09},\cite{Xue11}, \cite{Ueda08} and \cite{Cappelluti09} for the CDFS, C-COSMOS, SXDS and XMM-COSMOS sample, respectively; (2) right ascension and (3) declination (J2000) of the X-ray source; (4) adopted redshift; (5) 1: spectroscopic redshift, 2: photometric redshift; (6) redshift reference; 1: \cite{Civano12}, 2: zCOSMOS, 3: \cite{Vito13}, 4: \cite{Hiroi12}, 5: \cite{Brusa10}, 6: \cite{Salvato11}; (7) best-fitting column density in units of $[10^{22}\,\rmn{cm^{-2}}]$, errors are at the 90 per cent confidence limit; (8) soft-band flux in units of $[10^{-15} \,\rmn{erg\,cm^{-2}s^{-1}}]$; (9) intrinsic (i.e. absorption-corrected) rest-frame $2-10\,\rmn{keV}$ luminosity, in units of $10^{44}\,\rmn{erg\,s^{-1}}$. 
\end{table*}

\addtocounter{table}{-1}

\begin{table*} 
\caption{Continued.}
\begin{tabular}{|r|r|r|r|r|r|r|r|r|r|r|}
\hline
  \multicolumn{1}{c|}{ID} &
  \multicolumn{1}{c|}{RA} &
  \multicolumn{1}{c|}{DEC} &
  \multicolumn{1}{c|}{zadopt} &
  \multicolumn{1}{c|}{ztype} &
  \multicolumn{1}{c|}{zref} &
  \multicolumn{1}{c|}{$N_H$} &
  \multicolumn{1}{c|}{$F_{0.5-2\,\rmn{keV}} $} &
  \multicolumn{1}{c|}{$L_{2-10\,\rmn{keV}} $} \\
\hline
  \multicolumn{9}{c}{C-COSMOS}\\
  \hline
1658  & 150.28776 & 1.650687 & 3.871 & 2 & 1 & $<10$                & 1.64 & 3.05\\
1672  & 150.34429 & 1.635908 & 3.805 & 2 & 1 & $<67$                & 1.19 & 4.87\\
2059  & 150.31583 & 2.336873 & 4.216 & 2 & 1 & $127_{-99}^{+165}$   & 0.11 & 2.96\\
2220  & 149.78373 & 2.452045 & 5.07  & 1 & 1 & $<90$                & 0.80 & 4.68\\
2518  & 149.77105 & 2.365836 & 3.447 & 2 & 1 & $<118$               & 0.24 & 1.45\\
\smallskip
3293  & 150.30590 & 1.761653 & 3.265 & 2 & 1 & $<46$                & 0.38 & 0.96\\
3391  & 150.04273 & 1.872178 & 3.371 & 1 & 1 & $<12$                & 0.81 & 2.12\\
3397  & 150.06217 & 1.722708 & 3.033 & 2 & 1 & $<30$                & 0.96 & 2.02\\
3440  & 149.88939 & 1.966029 & 3.053 & 2 & 1 & $<28$                & 0.60 & 0.65\\
3449  & 149.91073 & 1.899629 & 3.063 & 2 & 1 & $-1$                 & 0.11 & 0.12\\
3636  & 150.13372 & 2.457497 & 3.189 & 1 & 1 & $25_{-22}^{+39}$     & 0.50 & 1.83\\
3651  & 150.17640 & 2.569708 & 3.144 & 2 & 1 & $<35$                & 0.50 & 0.58\\
  \hline
  \multicolumn{9}{c}{CDFS}\\
  \hline  
27    & 52.96054  & -27.87706& 4.385 & 2 & 3  & $68_{-29}^{+41}$     & 0.25 & 2.80\\
100   & 53.01658  & -27.74489& 3.877 & 2 & 3  & $38_{-24}^{+27}$     & 0.15 & 0.89\\
107   & 53.01975  & -27.66267& 3.808 & 2 & 3  & $<71$                & 0.48 & 1.52\\
132   & 53.03071  & -27.82836& 3.528 & 2 & 3  & $13_{-11}^{+16}$     & 0.11 & 0.31\\
170   & 53.04746  & -27.87047& 3.999 & 2 & 3  & $35_{-4}^{+4}$       & 0.16 & 9.43\\
  \smallskip
235   & 53.07029  & -27.84564& 3.712 & 1 & 3  & $<36$                & 0.04 & 0.08\\
262   & 53.07854  & -27.85992& 3.66  & 1 & 3  & $85_{-20}^{+22}$     & 0.12 & 1.63\\
283   & 53.08467  & -27.70811& 3.204 & 2 & 3  & $55_{-22}^{+39}$     & 0.10 & 0.80\\
285   & 53.08558  & -27.85822& 4.253 & 2 & 3  & $<6$                 & 0.06 & 0.12\\
331   & 53.10271  & -27.86061& 3.78  & 2 & 3  & $<23$                & 0.05 & 0.08\\
371   & 53.11158  & -27.76789& 3.101 & 2 & 3  & $35_{-10}^{+12}$     & 0.12 & 0.57\\
  \smallskip
386   & 53.11796  & -27.73439& 3.256 & 1 & 3  & $<7$                 & 0.09 & 0.11\\
412   & 53.12442  & -27.85169& 3.7   & 1 & 3  & $82_{-11}^{+12}$     & 0.21 & 2.85\\
458   & 53.13854  & -27.82128& 3.474 & 1 & 3  & $93_{-70}^{+74}$     & 0.02 & 0.26\\
521   & 53.15850  & -27.73372& 3.417 & 2 & 3  & $<23$                & 0.09 & 0.21\\
528   & 53.16158  & -27.85606& 3.951 & 1 & 3  & $74_{-24}^{+24}$     & 0.10 & 1.16\\
  \smallskip
546   & 53.16533  & -27.81419& 3.064 & 1 & 3  & $52_{-4}^{+4}$       & 0.67 & 4.73\\
556   & 53.17012  & -27.92975& 3.528 & 2 & 3  & $97_{-10}^{+9}$      & 0.76 & 11.40\\
563   & 53.17442  & -27.86742& 3.61  & 1 & 3  & $6_{-2}^{+2}$        & 2.07 & 4.45\\
573   & 53.17850  & -27.78411& 3.193 & 1 & 3  & $3_{-2}^{+2}$        & 0.81 & 1.19\\
588   & 53.18467  & -27.88103& 3.471 & 1 & 3  & $<3$                 & 0.65 & 0.84\\
  \smallskip
642   & 53.20821  & -27.74994& 3.769 & 2 & 3  & $<13$                & 0.09 & 0.15\\
645   & 53.20933  & -27.88119& 3.47  & 1 & 3  & $15_{-2}^{+2}$       & 3.16 & 9.24\\
651   & 53.21529  & -27.87033& 4.658 & 2 & 3  & $151_{-35}^{+39}$    & 0.13 & 4.64\\
674   & 53.24004  & -27.76361& 3.082 & 1 & 3  & $<7$                 & 0.70 & 0.86\\
700   & 53.26250  & -27.86308& 4.253 & 2 & 3  & $18_{-14}^{+16}$     & 0.74 & 3.11\\
717   & 53.28000  & -27.79892& 4.635 & 1 & 3  & $87_{-51}^{+65}$     & 0.12 & 1.85\\
\hline
    \multicolumn{9}{c}{SXDS}\\
  \hline  
16    & 33.93335  & -4.92384 & 3.512 & 1 & 4 & $<7$                 & 3.16 & 3.57\\
99    & 34.08598  & -5.28820 & 3.19  & 1 & 4 & $<1$                 & 4.68 & 4.19\\
154   & 34.14188  & -4.90640 & 3.6   & 2 & 4 & $4_{-3}^{+10}$        & 1.58 & 1.89\\
177   & 34.16076  & -5.17883 & 3.182 & 1 & 4 & $<4$                 & 3.35 & 2.98\\
179   & 34.16469  & -4.72107 & 3.426 & 2 & 4 & $<1$                 & 1.23 & 1.31\\
284   & 34.24511  & -4.81274 & 3.046 & 2 & 4 & $<3$                 & 2.44 & 1.96\\
287   & 34.24634  & -4.83036 & 4.09  & 2 & 4 & $<4$                 & 5.12 & 8.30\\
335   & 34.27461  & -5.22714 & 3.222 & 1 & 4 & $8_{-7}^{+13}$        & 2.60 & 2.39\\

  \hline\end{tabular}

\end{table*}

\addtocounter{table}{-1}

\begin{table*} 
\caption{Continued.}
\begin{tabular}{|r|r|r|r|r|r|r|r|r|r|r|}
\hline
  \multicolumn{1}{c|}{ID} &
  \multicolumn{1}{c|}{RA} &
  \multicolumn{1}{c|}{DEC} &
  \multicolumn{1}{c|}{zadopt} &
  \multicolumn{1}{c|}{ztype} &
  \multicolumn{1}{c|}{zref} &
  \multicolumn{1}{c|}{$N_H$} &
  \multicolumn{1}{c|}{$F_{0.5-2\,\rmn{keV}} $} &
  \multicolumn{1}{c|}{$L_{2-10\,\rmn{keV}} $} \\
\hline
    \multicolumn{9}{c}{SXDS}\\
    \hline
342   & 34.28106  & -4.56844 & 3.048 & 2 & 4 & $25_{-19}^{+120}$    & 6.57 & 5.28\\
385   & 34.32280  & -5.08699 & 3.39  & 2 & 4 & $<8         $        & 2.66 & 2.76\\
422   & 34.34430  & -5.39438 & 3.422 & 1 & 4 & $3_{-2}^{+2}$        & 6.02 & 6.38\\
449   & 34.35595  & -4.98441 & 3.328 & 1 & 4 & $<4$                 & 2.58 & 2.56\\
459   & 34.36505  & -5.28871 & 3.969 & 1 & 4 & $<1$                 & 1.60 & 2.42\\
    \smallskip
489   & 34.38058  & -5.09214 & 3.334 & 2 & 4 & $7_{-4}^{+5}$        & 4.65 & 4.63\\
508   & 34.39336  & -5.08771 & 3.975 & 1 & 4 & $<7$        & 2.27 & 3.45\\
520   & 34.39922  & -4.70872 & 3.292 & 1 & 4 & $<15$       & 1.57 & 1.52\\
564   & 34.43107  & -4.54404 & 3.204 & 1 & 4 & $<6$                 & 3.01 & 2.72\\
650   & 34.49477  & -5.13762 & 3.032 & 1 & 4 & $<1$                 & 2.18 & 1.73\\
  \smallskip
700   & 34.52910  & -5.06942 & 3.128 & 1 & 4 & $5_{-2}^{+2}$        & 7.22 & 6.18\\
742   & 34.56485  & -5.40081 & 3.114 & 1 & 4 & $<2$                 & 6.84 & 5.78\\
788   & 34.59990  & -5.10029 & 4.096 & 2 & 4 & $<6$                 & 2.07 & 3.37\\
809   & 34.61807  & -5.26411 & 3.857 & 1 & 4 & $<6$                 & 1.96 & 2.76\\
824   & 34.63105  & -4.73291 & 3.699 & 1 & 4 & $<18$                & 2.95 & 3.77\\
835   & 34.64068  & -5.28748 & 3.553 & 1 & 4 & $<3$                 & 1.34 & 1.55\\
  \smallskip
888   & 34.68521  & -4.80691 & 4.55  & 1 & 4 & $<17$                & 3.72 & 7.76\\
904   & 34.69847  & -5.38866 & 3.02  & 1 & 4 & $<1$                 &21.60 & 1.69\\
926   & 34.72080  & -5.01810 & 3.264 & 1 & 4 & $10_{-6}^{+11}$       & 1.09 & 10.40\\
930   & 34.72563  & -5.52342 & 3.49  & 2 & 4 & $<4$                 & 4.67 & 5.19\\
1032  & 34.80911  & -5.17238 & 3.584 & 2 & 4 & $<6$                 & 4.19 & 4.96\\
1238  & 35.09164  & -5.07500 & 4.174 & 1 & 4 & $<2$                 & 2.60 & 4.42\\
\hline
    \multicolumn{9}{c}{XMM-COSMOS}\\
    \hline
187   & 150.240869& 2.658730 & 3.356 & 1 & 5 &                      & 3.80 & 4.85\\
326   & 150.256847& 2.646315 & 3.003 & 1 & 5 &                      & 1.20 & 1.19\\
2421  & 149.528973& 2.380177 & 3.097 & 1 & 5 &                      & 5.20 & 5.53\\
2518  & 150.489739& 1.746145 & 3.176 & 1 & 5 &                      & 1.30 & 1.46\\
5116  & 150.735835& 2.199860 & 3.5   & 1 & 1 &                      & 4.10 & 5.75\\
5120  & 149.761660& 2.434940 & 3.647 & 2 & 6 &                      & 2.50 & 3.84\\
  \smallskip
5161  & 149.748141& 2.732395 & 3.169 & 2 & 5 &                      & 1.60 & 1.79\\
5162  & 149.755002& 2.738672 & 3.524 & 1 & 5 &                      & 1.90 & 2.70\\
5175  & 150.620110& 2.671575 & 3.143 & 1 & 5 &                      & 6.60 & 7.26\\
5199  & 149.471980& 2.793812 & 3.626 & 1 & 5 &                      & 2.70 & 4.10\\
5219  & 150.736810& 2.722455 & 3.302 & 1 & 5 &                      & 2.20 & 2.70\\
5259  & 150.466760& 2.531320 & 4.45  & 2 & 5 &                      & 1.10 & 2.63\\
  \smallskip
5331  & 150.608403& 2.769818 & 3.038 & 1 & 5 &                      & 4.60 & 4.68\\
5345  & 150.584647& 2.081176 & 3.296 & 2 & 6 &                      & 1.90 & 2.33\\
5347  & 149.669389& 2.167760 & 3.089 & 1 & 1 &                      & 2.00 & 2.11\\
5382  & 150.440006& 2.703517 & 3.465 & 1 & 5 &                      & 1.10 & 1.51\\
5592  & 150.703941& 2.369606 & 3.749 & 1 & 5 &                      & 2.20 & 3.60\\
5594  & 149.467412& 1.855175 & 4.161 & 1 & 5 &                      & 1.30 & 2.68\\
  \smallskip
5606  & 149.776790& 2.444055 & 4.166 & 1 & 5 &                      & 1.30 & 2.69\\
10690 & 150.596966& 2.432707 & 3.1   & 2 & 5 &                      & 1.30 & 1.39\\
54161 & 149.707096& 2.525978 & 3.003 & 2 & 6 &                      & 0.80 & 0.79\\
60017 & 150.112295& 2.845933 & 3.062 & 2 & 6 &                      & 1.30 & 1.35\\
60311 & 149.419259& 2.883111 & 3.329 & 2 & 6 &                      & 3.00 & 3.76\\
60391 & 149.467610& 2.531345 & 3.518 & 2 & 6 &                      & 1.00 & 1.42\\
60465 & 149.532828& 1.958666 & 3.146 & 2 & 6 &                      & 1.00 & 1.10\\

  \hline\end{tabular}

\end{table*}

\subsection{General properties of the sample}

\begin{figure}
\includegraphics[width=80mm,keepaspectratio]{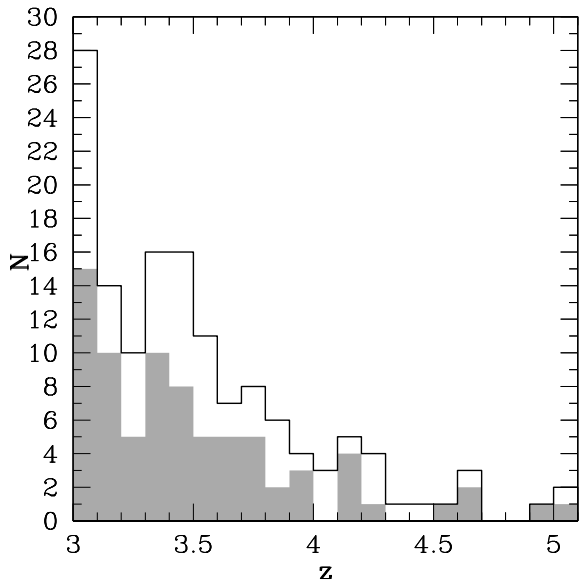} 
\caption{Redshift distribution of the 141 sources in the redshift range $3<z<5.1$ (black histogram) and the fraction of sources with spectroscopic redshift (grey shaded histogram) }
\label{fig1}
 \end{figure}

\begin{figure}
\includegraphics[width=80mm,keepaspectratio]{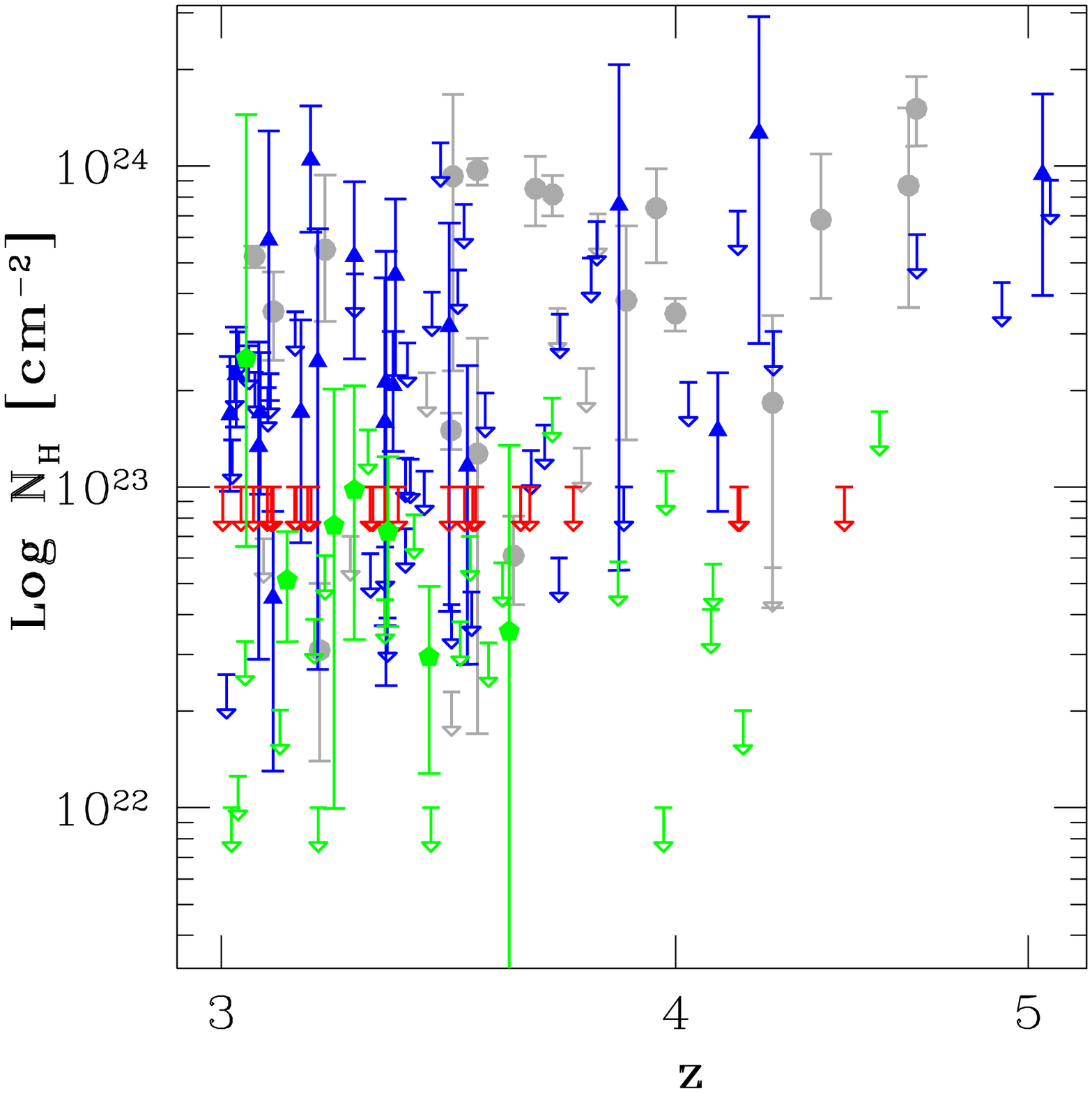} 
\includegraphics[width=80mm,keepaspectratio]{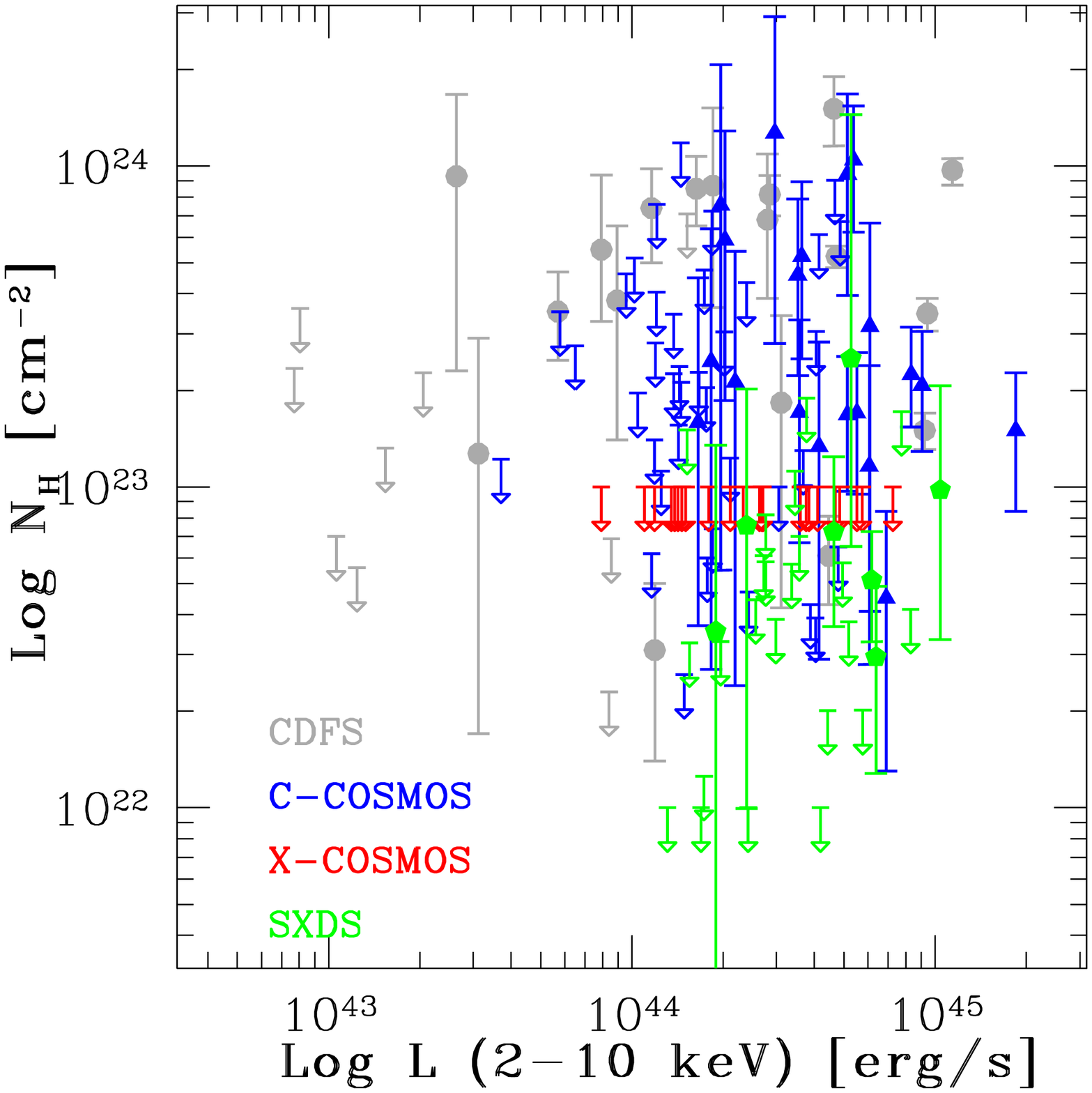}
\caption{Column density as a function of redshift (upper panel) and rest-frame, intrinsic hard-band luminosity (lower panel) for the 141 sources at $3<z<5.1$. Grey, blue, red and green symbols represent sources from the CDFS, C-COSMOS, XMM-COSMOS and SXDS sample, respectively. Errors are plotted at the 90 per cent confidence level (c.l.). XMM-COSMOS sources are assumed to be unobscured and are plotted as upper limits corresponding to $N_H<10^{23}\,\rmn{cm^{-2}}$.}
\label{fig2}
 \end{figure}

Fig.~\ref{fig1} shows the redshift distribution of the sample.
The column density is plotted against redshift (upper panel) and luminosity (lower panel) for each source of the sample in Fig.~\ref{fig2}. We defined the column density to be constrained if its lower limit at the 90 per cent confidence level (c.l.) is larger than zero, otherwise we plot its 90 per cent c.l.  upper limit as a downward pointing arrow. Since in \cite{Hiroi12} errors are quoted at the $1\sigma$ c.l., for sources detected by the SXDS survey we applied the same definition, but changed the confidence limit to the 90 per cent c.l. Thirty-six out of the 141 objects are obscured by a column density $N_H>10^{23}\,\rmn{cm^{-2}}$. Most of them are detected by the CDFS and C-COSMOS survey, as expected, since they are the two deepest surveys among those considered in this work.

\begin{figure}
\includegraphics[width=80mm,keepaspectratio]{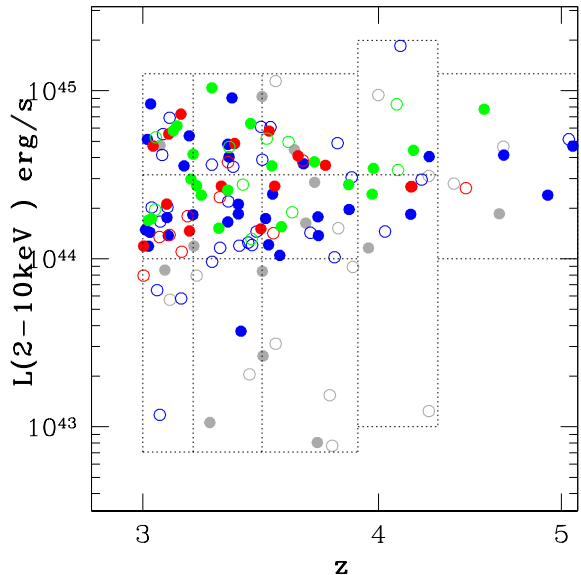}
\caption{Rest-frame, intrinsic hard-band luminosity as a function of redshift. The colour code is the same as in Fig. \ref{fig2}. Filled and empty circles refer to sources with spectroscopic and photometric redshift, respectively.} The dashed rectangles represent the luminosity-redshift bins considered in \S~\ref{method}.
\label{fig3}
 \end{figure}

 Fig.~\ref{fig3} shows the luminosity plotted against the redshift. The importance of collecting samples from different (in terms of deepness and area) surveys to sample different region in the $L_X-z-N_H$ space can be inferred from these figures.

\section{The Hard X-Ray Luminosity Function at $3<z<5.1$}
\label{HXLF}
The differential luminosity function of any population of extra-galactic objects can be defined as the number $N$ of objects per unit comoving volume $V$ per unit logarithmic luminosity $\rmn{logL_X}$
\begin{equation}\label{phi}
 \phi=\frac{\rmn{d}\Phi}{\rmn{dlogL_X}}(z,\rmn{logL_X}) = \frac{\rmn{d^2}N}{\rmn{d}V\rmn{dlogL_X}}
\end{equation}

In \S~\ref{method} we describe the method we used to compute the binned hard band ($2-10\,\rmn{keV}$) luminosity function (HXLF). In \S~\ref{HXLFfit} we present the analytical models assumed to fit the HXLF and the results of the fit, with no correction for redshift incompleteness. In \S~\ref{complete} we introduce this correction in the fitting procedure.

\subsection{Method}\label{method}

A binned representation of the HXLF can be derived using the $1/V_{max}$ method (\citealt{Schmidt68,Avni80}), but it suffers from systematic bias for objects close to the flux limit of the surveys (\citealt{Page00}). Since this is the case of most sources included in our high-redshift sample, we preferred to use the \cite{Page00} method, which was developed to fix this problem. Therefore, our estimation of the binned HXLF in a given redshift ($\Delta z$) and luminosity ($\Delta \rmn{logL_X}$) bin has the following form:

\begin{equation}\label{PageHXLF}
\phi=\frac{N}{\int\!\!\int \Omega\Theta\frac{\rmn{d}V}{\rmn{d}z}\,\rmn{d}z\,\rmn{dlogL_X}}
\end{equation}

where $N$ is the number of sources in the bin $ \Delta \rmn{logL_X}-\Delta z$, $\Omega=\Omega(z,L_X,N_H)$ is the sky coverage (i.e. the fraction of the sky covered by a survey at a given soft-band flux) and $\Theta = \Theta(z,L_X,N_H)$ is a factor which accounts for the redshift incompleteness of the parent sample.

The \cite{Page00} method assumes that the luminosity function variation inside a bin $\Delta \rmn{logL_X}-\Delta z $ is small enough that can be neglected. This assumption is not necessarily true in our case, since we will make use of relatively large bins of redshift and luminosity. However, this would not affect the fit parameters which will be derived in \S~\ref{HXLFfit}, since the fit will be performed on unbinned data. Moreover, this method provides a formally correct derivation of the errors on the binned HXLF, while the $1/V_{max}$ method does not.

\subsubsection{Sky coverage and absorption}\label{skycov}
The $0.5-2\,\rmn{keV}$ sky coverage of the CDFS survey was taken from \cite{Xue11}. As for the SXDS survey, we used the sky coverage of the overlapping region between the X-ray and optical surveys (Ueda Y., private communication), used also by \cite{Hiroi12} and derived from the curve presented in \cite{Ueda08}. The C-COSMOS sky coverage was published by \cite{Elvis09}. Since we cut out an external frame of this field (see \S~\ref{C-COSMOS}), a region of low-medium exposure, we imposed the maximum area of the survey to be $0.72\,\rmn{deg^2}$, the area of the reduced C-COSMOS field, without any change in the normalization and shape of the sky coverage at low fluxes. As for the reduced XMM-COSMOS field, we re-computed the sky coverage as in \cite{Cappelluti09}, excluding the central $0.72\,\rmn{deg^2}$ region (see \S~\ref{XMM-COSMOS}).

Two main issues arise when considering the sky coverage curves: 1) the sky coverage for each survey was computed assuming a different spectral shape (\citealt{Xue11,Elvis09,Cappelluti09,Ueda08}); 2) sources with a given hard-band, intrinsic luminosity, have different soft-band fluxes and spectral shape, because of the effect of absorption. 
To address these issues, we converted the sky coverage as a function of flux of the $i$-th survey $\Omega_i(F)$ into a sky coverage as a function of count rate $\Omega_i(CR)$, according to the specific spectral shape assumed for each reference paper. Therefore, the products are no more dependent on a specific spectral shape.

Then, the full sample was splitted between two subsamples: sources with a column density (column 8 of Tab.~\ref{tab1})  $\rmn{logN_H}>23$ were included in the subsample A (absorbed; 36 objects), the others were counted in the subsample U (unabsorbed; 105 objects). We used this threshold value because it is approximately the $N_H$ detection limit at $z \ga 3$: lower values of column density do not affect significantly the observed soft-band spectrum of a X-ray source, since the photoelectric cut-off shifts at lower energies (e.g. \citealt{Vito13}) and, hence, cannot be constrained. 

The $\Omega_i(CR)$ are re-converted into sky coverage as a function of flux assuming as spectral shape a simple power-law with spectral index $\Gamma=1.8$ and no absorption for the subsample U ($\Omega_i^{\rmn{U}}$) and accounting for an absorption due to a column density $\rmn{logN_H}=23.5$ at $z=3.5$ (approximatively the median values of $N_H$ and redshift for obscured sources, see Tab.~\ref{tab1}) for the subsample A ($\Omega_i^{\rmn{A}}$). 
 The use of the median $N_H$ and $z$ is justified by the need of a single spectral shape during the computation of the binned HXLF and by the large uncertainties on the column density of each individual source.
All luminosities for the CDFS, C-COSMOS and XMM-COSMOS samples were computed with the above photon index.
The luminosity for the SXDS sample were obtained by \cite{Hiroi12} assuming a power-law with $\Gamma=1.9$ plus a reflection component. However, this difference in the spectral shape is completely negligible for the final results.

In this procedure, we used typical (i.e. on axis) response matrices of the CDFS, C-COSMOS and XMM-COSMOS surveys to compute the energy-to-count-rate conversion factors (ECF) for the relative samples. The adopted matrices do not affect significantly the final results (using different responses resulted in a difference of $<5$ per cent), since the conversion is made twice in the reverse order (from flux to count rate and then from count rate to flux) with the same matrix. As for the SXDS survey, we used the ECF provided by \cite{Ueda08} for different values of effective photon index. We could use the ECF corresponding to $\Gamma=1.8$ for the U subsample. For the A subsample an effective photon index $\Gamma\sim-1$ in the soft band was derived from an intrinsic $\Gamma=1.8$ power-law spectrum absorbed by a column density $\rmn{logN_H}=23.5$ at $z=3.5$. Since \cite{Ueda08} did not provide the ECF corresponding to $\Gamma=-1$, we linearly interpolated the values they reported and derived the ECF. 
dependence of

\begin{figure}
\includegraphics[width=80mm,keepaspectratio]{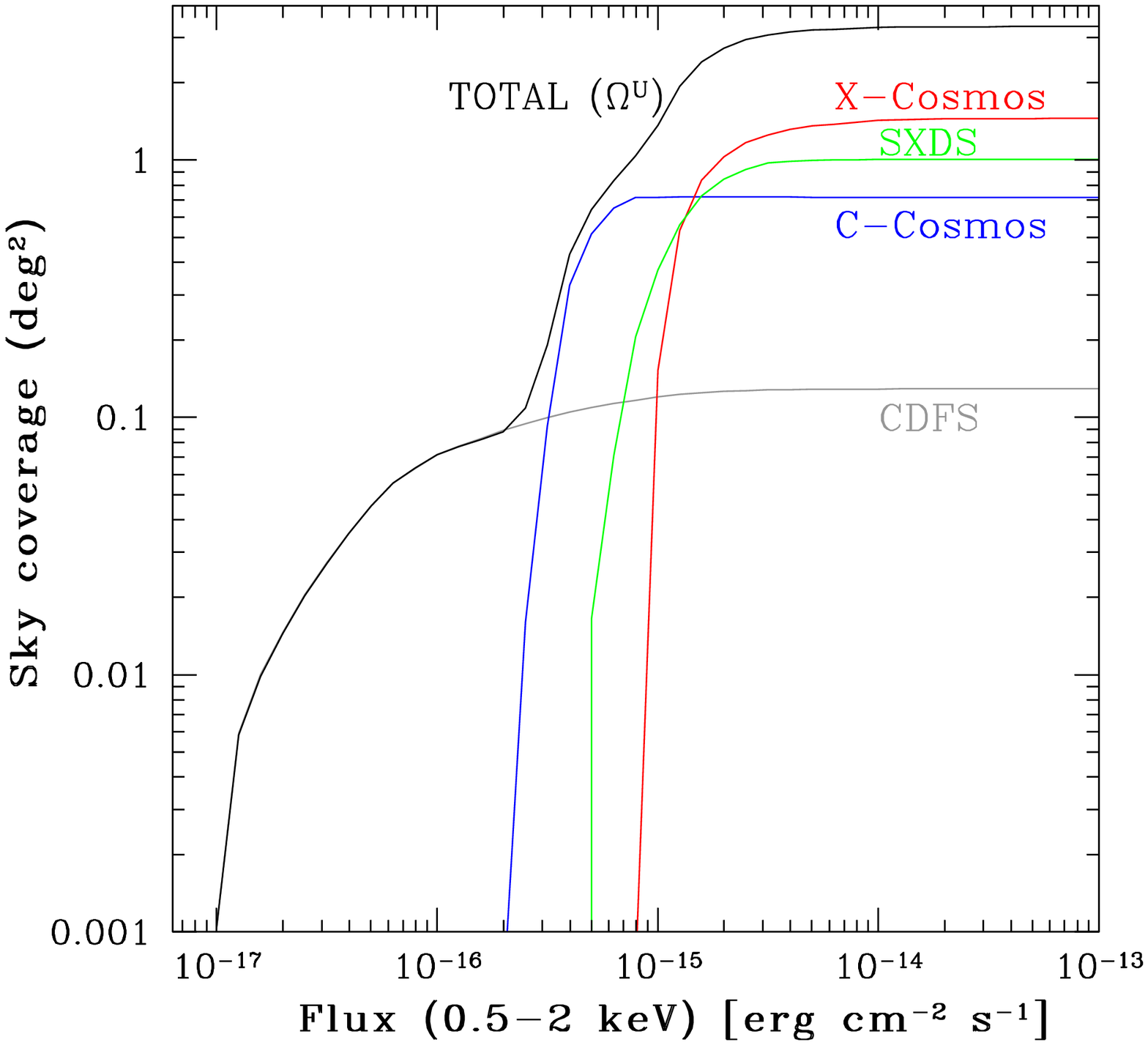}
\includegraphics[width=80mm,keepaspectratio]{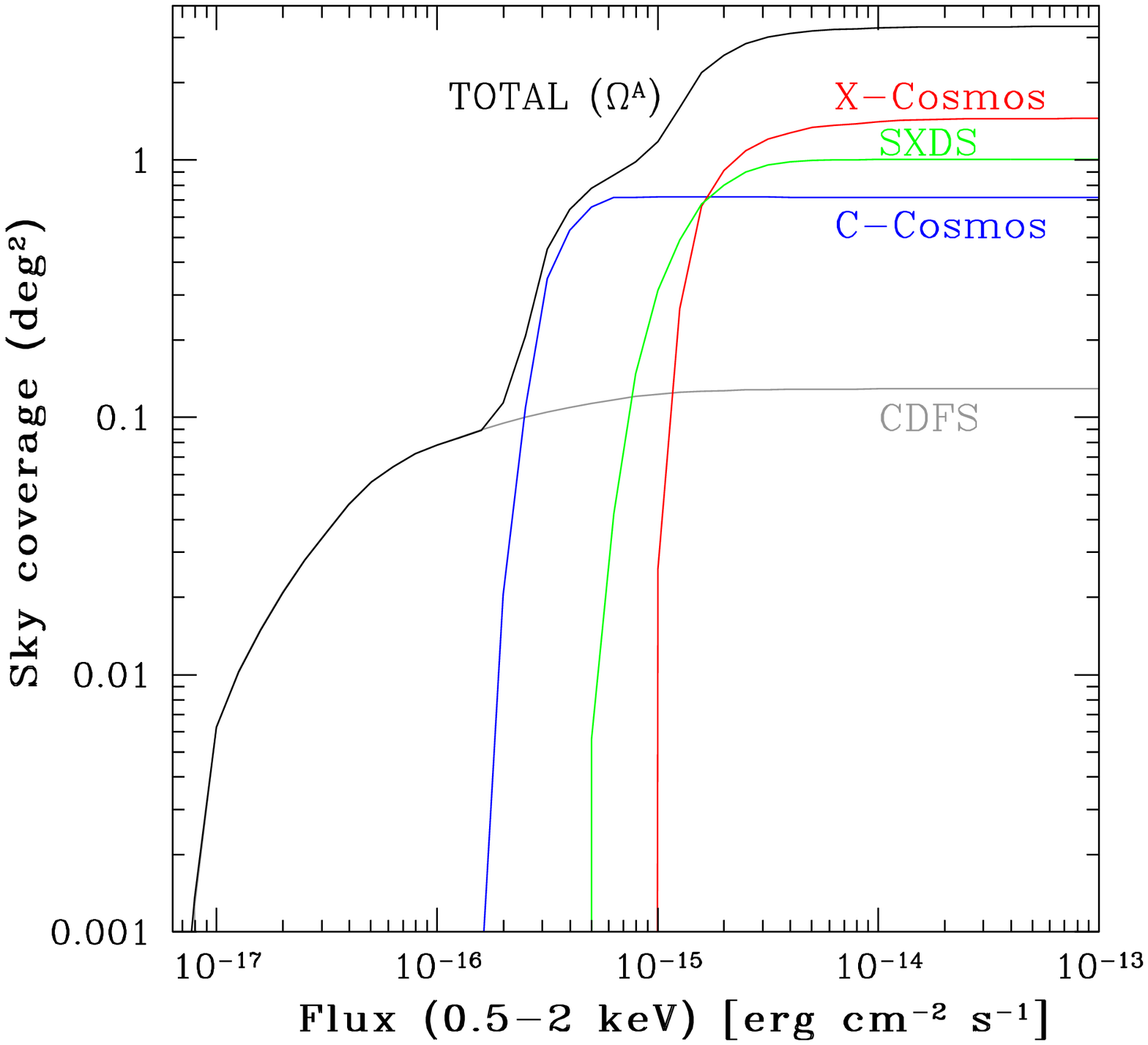}
\caption{Sky coverage used for the computation of the HXLF for the unabsorbed (\textit{upper panel}) and absorbed (\textit{lower panel}) source sample. The total sky coverage is computed as the sum of the individual surveys sky coverage, as described in \S~\ref{skycov}.}
\label{fig4}
 \end{figure}

We computed separately the HXLF for the subsamples U ($\phi^{\rmn{U}}$) and A ($\phi^{\rmn{A}}$). The sky coverage $\Omega^{\rmn{U}}=\sum\nolimits_i\Omega_i^{\rmn{U}}$ (Fig.~\ref{fig4}, upper panel) is used to compute $\phi^{\rmn{U}}$ and, similarly, $\Omega^{\rmn{A}}=\sum\nolimits_i\Omega_i^{\rmn{A}}$ (Fig.~\ref{fig4}, lower panel) to compute $\phi^{\rmn{A}}$.
Using this formalism, we adopted the coherent addition of samples in the sense given by \citet[ i.e. the proper addition of samples from different flux-limited surveys, assuming that each object could in principle be detected in all of them]{Avni80} and already used in a similar way by e.g. \cite{Yencho09}. Though the curves in Fig.~\ref{fig4} appear very similar, we note that close to the flux limit of a survey, the coverages for the U and A subsamples can be significantly different (e.g., a factor of $\sim4$ for C-COSMOS at $F_{0.5-2\,\rmn{keV}}=3\times10^{-16}\rmn{erg}\,\rmn{cm^{-2}}\rmn{s^{-1}}$).

For a given point in the $\Delta\rmn{log} L_X -\Delta z$ space, the soft band flux, required to obtain the sky coverage, is derived from the intrinsic hard band luminosity assuming a simple power-law with $\Gamma=1.8$ for the U subsample, and accounting also for an absorption due to a column density of $N_H=10^{23.5}\rmn{cm^{-2}}$ at that redshift for the A subsample.

Finally, we computed the total HXLF as the sum of the HXLF of the two subsamples:

\begin{equation}\label{HXLFsum}
\phi^{\rmn{TOT}} = \phi^{\rmn{U}} + \phi^{\rmn{A}}
\end{equation}

Errors are computed as the $1\sigma$ Poissonian uncertainties \citep{Gehrels86} for $\phi^{\rmn{U}}$ and $\phi^{\rmn{A}}$, and then propagated to obtain the errors on $\phi^{\rmn{TOT}}$.

 We computed the binned HXLF in 12 bin of the $\Delta\rmn{log} L_X-\Delta z$ plane (Fig.~\ref{fig3} and Tab.~\ref{zHL}) using Eq. \ref{HXLFsum}. Bins were chosen to have an acceptable statistics in each of them: we required at least 5 sources per bin, at least one of which obscured, and a minimum bin size of 0.5 dex in luminosity to increase the statistics in the most populated regions of the $\Delta\rmn{log} L_X-\Delta z$ plane. The resulting HXLF is shown in Fig.~\ref{fig5}. At this stage, we did not include any correction for the redshift incompleteness. 

\begin{table*}

\caption{Number of objects in each $\Delta \rmn{logL_X} - \Delta z$ bin. Number of obscured objects are between brackets.} 
\begin{tabular}{r|rrrrr}
\hline
  \multicolumn{1}{r|}{$\Delta \rmn{LogL_X}$} &
  \multicolumn{1}{r}{{\bf z=3-3.19}} &
  \multicolumn{1}{r}{{\bf3.19-3.47}} &
    \multicolumn{1}{r}{{\bf3.47-3.9}} &
   \multicolumn{1}{r}{{\bf3.9-4.3}} & 
  \multicolumn{1}{r}{{\bf4.3-5.1}} \\
  \cline{2-6}
  {\bf  42.85-44}   &  6 (1)  &  5 (1)  & 7 (3)  &  ---    &   ---    \\
  {\bf  43-44.5}    &  ---    &  ---    & ---    &  9 (3)  &   ---    \\
  {\bf  44-44.5}    &  20 (2) &  19 (2) & 18 (3) &  ---    &   ---    \\
  {\bf  44-45.1}    &  ---    &  ---    & ---    &  ---    &   9 (4)  \\
  {\bf  44.5-45.1}  &  14 (8) &  13 (5) & 14 (2) &  ---    &   ---    \\
  {\bf  44.5-45.3}  &  ---    &  ---    & ---    &  7 (2)  &   ---    \\

  \hline
\end{tabular}

\label{zHL}
\end{table*}

 \subsection{Fit to the HXLF}\label{HXLFfit}
 \subsubsection{Models}\label{models}
In order to derive an analytical representation of the HXLF, we assumed different models, starting with a non-evolving, smoothed double power-law:

\begin{equation}\label{D-PL}
 \phi=\frac{\rmn{d}\Phi(L_X)}{\rmn{dLog}L}=A\Big[ \Big( \frac{L_X}{L_*}\Big)^{\gamma_1}  + \Big( \frac{L_X}{L_*}\Big)^{\gamma_2}\Big]^{-1}
\end{equation}

where $A$ is a normalization factor, $\gamma_1$ and $\gamma_2$ are the slopes of the faint end bright end, respectively, and $L_*$ is the break luminosity \citep{Miyaji00,Hasinger05}.

We then introduced an evolution with redshift. First, we included a pure luminosity evolution (PLE) factor $p_{lum}$, which preserves the shape of $\phi$ but shifts it on the luminosity axis. The analytical expression is the same as Eq. \ref{D-PL} but $L_*$ is multiplied by 

\begin{equation}\label{PLE_eq}
 e_{lum}(z)=(\frac{1+z}{1+z_{min}})^{p_{lum}}
\end{equation}

where $p_{lum}$ is the luminosity evolution factor and $z_{min} = 3$. 
Then, we investigated the case of a pure density evolution (PDE), in which the normalization of the HXLF is a function of redshift, multiplying Eq. \ref{D-PL} by a factor

\begin{equation}\label{PDE_eq}
 e_{den}(z) = (\frac{1+z}{1+z_{min}})^{p_{den}}
\end{equation}
where $p_{den}$ is the density evolution factor.

We also accounted for an independent luminosity and density evolution (ILDE, \citealt{Yencho09}), which acts independently on the luminosity and normalization.  
Therefore, both the evolutionary factors in Eq. \ref{PLE_eq} and \ref{PDE_eq} were applied simultaneously on Eq. \ref{D-PL}.

\cite{Aird10} proposed a luminosity and density evolution model (LADE) similar to the ILDE model, except for a different parametrization of the density evolution and a more complex luminosity evolution, which is assumed to experience a smooth transition between two different regime at a critical redshift. We fitted the LADE model but have slightly changed the analytical form. Since the best-fitting critical redshift found by \citet[$z_c=0.75\pm0.09$]{Aird10} is far from the redshift range probed by this work, we assumed a luminosity evolution with a single slope, parametrized by Eq.~\ref{PLE_eq}. As for the density evolution, we multiplied the normalization $A$ by

\begin{equation}\label{LADE_eq}
 e_{den}(z) = \frac{10^{p_{den}(1+z)}}{10^{p_{den}(1+z_{min})}}.
\end{equation}
In this form, the parameter $A$ refers to the normalization at $z=3$, as for the other models.

 Finally, we assumed a luminosity-dependent density evolution (LDDE, \citealt{Schmidt83}) model. In this case, following the parametrization by \cite{Hasinger05}, Eq.~\ref{D-PL} is multiplied by 

\begin{equation}\label{LDDE_eq}
 e_{den}(z,L) = (\frac{1+z}{1+z_{min}})^{p_{den} + \beta (\rmn{logL}-44)}
\end{equation}

where $\beta$ is the parameter that accounts for the luminosity dependency. In this case, unlike the previous ones, the shape of the HXLF changes with redshift.

\subsubsection{Fit and results}\label{maxlik}
 
We evaluated the best-fit parameters for the different models using the unbinned maximum-likelihood method \citep{Marshall83}. The best-fitting parameters for each analytical model $\phi(z,L_X)$ are those which minimize the expression

\begin{equation}
 \mathcal{L}=-2\sum\limits_{i=1}\limits^{N}\rmn{ln}[\phi(z_i,L_i)] +2\!\!\int\!\!\!\!\!\int\!\phi(z,L)\Omega\Theta\frac{\rmn{d}V}{\rmn{d}z}\rmn{d}z\rmn{d}L
\end{equation}
where $N$ is the total number of sources, and the double integral is computed over the entire $\Delta\rmn{log} L_X - \Delta z$ region ($3<z\leq5.1$ and $42.85<\rmn{Log}L_X<45.3$).
Minimization and error analysis were performed with the MINUIT library~\footnote{http://lcgapp.cern.ch/project/cls/work-packages/mathlibs/minuit/index.html}. Confidence regions are computed at the $1\sigma$ c.l. by varying each parameter around the best-fitting value, leaving all the others parameters free, until $\Delta \mathcal{L} =1$. 

Since the sky coverage is different for the U and A subsamples, we minimized $\mathcal{L}^{TOT}=\mathcal{L}^{U}+\mathcal{L}^{A}$, where $\Omega^U$ and $\Omega^A$ are used for the respective subsamples. This is equivalent to perform a simultaneous fit to the two subsamples.  

\begin{figure*}
\includegraphics[width=160mm,keepaspectratio]{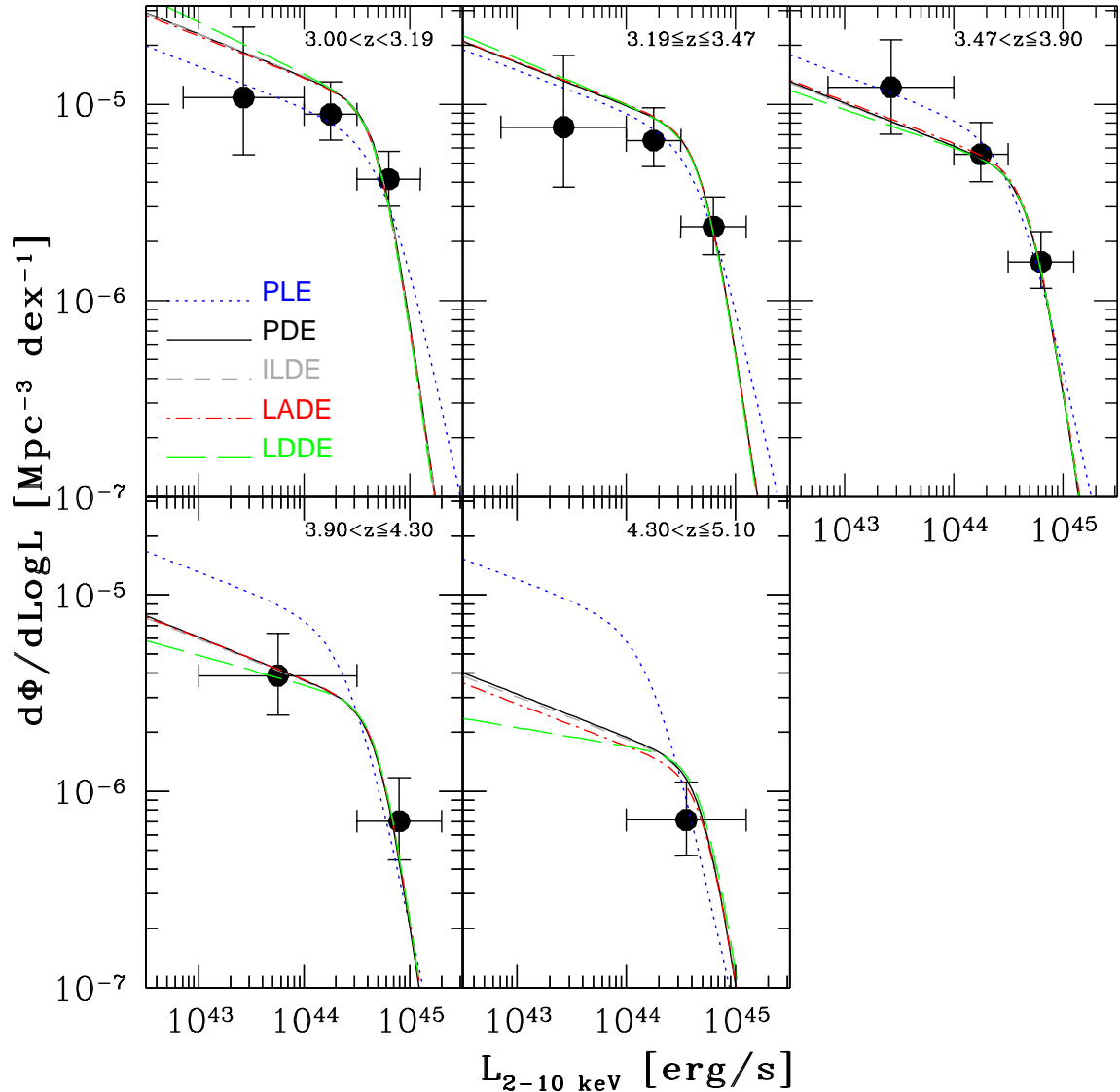}
\caption{ Binned luminosity function. The best fit for the models considered in \S~\ref{HXLFfit} are evaluated at the mean redshift of each bin and  plotted with different line styles and colours. The best-fitting ILDE model (short-dashed line) is barely visible over the PDE model (solid line), having very similar best-fitting parameters (see Tab.~\ref{tab2} and text).  No correction for redshift incompleteness has been included. } \label{fig5}

 \end{figure*}

We determined the normalization parameters $A^U$ and $A^A$ by imposing that the total number of sources is the observed one, separately for the U and A subsamples.  Their sum gives the normalization $A$ of the total HXLF. Errors on the normalization are computed separately for the two subsamples, accounting for the statistical errors on the number of sources, and then propagated to derive the error on $A$. Best-fitting models are plotted in Fig.~\ref{fig5} and the parameters are reported in Tab.~\ref{tab2}.

Unlike the $\chi^2$ test, the maximum likelihood method gives no information about the goodness of the fit. We evaluated it by using the bidimensional Kolmogorov-Smirnov test (2DKS; \citealt{Peacock83, Fasano87}). We mention that a value of the 2DKS probability $\gtrsim 0.2$ may not be accurate, but still means that the data and the model predictions are not significantly different \citep{Press92}. Also, the formal KS test derivation requires the two compared samples to be independent. This is not the case, since we will compare data and models derived by fitting the data themselves. Therefore, the resulting 2DKS test values are expected to be overestimated. However, following e.g. \cite{Miyaji00}, we chose to use these probabilities as goodness-of-fit indicators, as we are more interested in comparing different models rather than in the absolute probabilities, and not to apply the formally-correct, but computationally very expensive, treatment, which would involve large sets of Monte-Carlo 
simulations.

\begin{table*}
\caption{ Best-fit parameters.}\label{tab2}
\begin{tabular}{|r|r|r|r|r|r|r|r|r|}
\hline
  \multicolumn{1}{|r|}{ MODEL } &
  \multicolumn{1}{|r|}{{\bf $A $ }} &
  \multicolumn{1}{r|}{{\bf $L_*$}} &
  \multicolumn{1}{r|}{{\bf $\gamma_1$}} &
  \multicolumn{1}{r|}{{\bf $\gamma_2$}} &
  \multicolumn{1}{r|}{{\bf $p_{lum}$}} &
  \multicolumn{1}{r|}{{\bf $p_{den}$}} &
    \multicolumn{1}{r|}{{\bf $\beta$}} &
    \multicolumn{1}{r|}{ 2DKS}        \\
      (1)   &  (2)   & (3)  & (4)  &  (5)  &  (6)    & (7)    & (8) & (9)  \\

  \hline
  PLE  &  $0.65^{+0.06}_{-0.06}$  & $6.56_{-2.22}^{+2.38}$ &  $0.21_{-0.20}^{+0.16}$  &  $2.58_{-0.60}^{+0.75}$  &  $-3.73_{-0.92}^{+0.77}$  & ---    &---  & $0.05$   \\
  PDE  &  $1.10^{+0.11}_{-0.11}$  & $5.26_{-1.20}^{+1.06}$ &  $0.22_{-0.16}^{+0.13}$  &  $3.79_{-0.87}^{+1.08}$  &  ---    & $-6.00_{-0.87}^{+0.84}$  &---  & $0.38$   \\
  ILDE &  $1.13^{+0.11}_{-0.11}$  & $5.13_{-1.53}^{+1.32}$  &  $0.21_{-0.17}^{+0.13}$  &  $3.75_{-0.91}^{+1.10}$  &  $0.13_{-0.81}^{+0.94}$   & $-6.13_{-1.23}^{+1.17} $ &---  & $0.38$   \\
    LADE &  $1.08^{+0.11}_{-0.11}$  & $5.10_{-1.54}^{+1.33}$  &  $0.21_{-0.18}^{+0.13}$  &  $3.74_{-0.91}^{+1.11}$  &  $0.16_{-0.82}^{+0.95}$   & $-0.57_{-0.11}^{+0.11} $ &---  & $0.46$   \\
  LDDE & $1.05^{+0.10}_{-0.10}$   & $5.24_{-1.19}^{+1.05}$ & $0.28_{-0.19}^{+0.16}$  & $3.87_{-0.88}^{+1.08}$  & --- & $-6.43_{-1.17}^{+1.12}$ & $1.18_{-2.00}^{+2.06}$ & $0.42$ \\

  \hline

\end{tabular}

(1) model; (2) normalization in units of $10^{-5} \rmn{Mpc^{-3}}$; (3) knee luminosity in units of $10^{44}\rmn{erg}\,\rmn{s^{-1}}$; (4) faint and (5) bright end slope; (6) luminosity and (7) density evolutionary factor; (8) luminosity-dependency factor of the density evolution; (9) two dimensional Kolmogorov-Smirnov test probability.
\end{table*}

\subsection{Introducing the completeness factor}\label{complete}
So far we have not considered the redshift incompleteness of the sample and computed the HXLF assuming $\Theta=1$ in Eq. \ref{PageHXLF}. We then searched for a reasonable method to estimate it. A procedure adopted by some authors \citep[e.g.][]{Barger05,Yencho09} to provide upper limits to the complete HXLF is to count all sources with no redshift information in each $\Delta \rmn{logL_X} - \Delta z$ bin. However, this method would result in very loose constraints on the HXLF. Other authors developed procedures based on optical luminosity \citep[e.g.][]{Matute06}.

There are 31, 28 and 6 soft-band detected sources in the CDFS, C-Comsos and XMM-COSMOS surveys which do not have a spectroscopic or photometric redshift. \cite{Hiroi12} reported that for 6 AGN candidates in the SXDS neither a spectroscopic nor a photometric redshift was available. Since the catalogue of the optical survey in this region is not public, we do not know which are those 6 X-ray sources and we exclude them from the computation of the completeness correction. We do not expect significant changes if they would be included, since they represent a tiny fraction of the overall sample.

 We made the extreme assumption that all the 65 sources are at $3<z<5.1$ and applied the following method to derive $\Theta(z,L_X,N_H)$, i.e. the fraction of sources in the high-redshift sample with redshift information in a given $\Delta\rmn{log} L_X-\Delta z$ bin:
 each source with no redshift information, but for which the flux $F$ is known, is counted simultaneously in the A and U subsample with a weight ($w_A(F)$ and $w_U(F)=1-w_A(F)$, respectively) corresponding to the fraction of absorbed ($w_A(F)$) or unabsorbed ($w_U(F)$) sources with redshift at similar fluxes. Each weighted source is then spread in redshift accordingly to the normalized redshift distribution of the sources with redshift in the relative subsample ($P_A(z)$ and $P_U(z)$). The redshift distribution were evaluated in the same bins used for the binned HXLF. Different choices of bins do not result in significant change of the results and, anyhow, have the same level of arbitrariness. The luminosity was then computed for all the weighted sources in the two subsamples in each redshift bin, at the median redshift of the bin. Finally, knowing the number of sources with redshift and having estimated the number of those without in each $\Delta\rmn{log} L_X-\Delta z$ bin, we derived $\Theta(z,L_X,N_H)$. 
We note 
that using this 
procedure all the 65 sources are 
included, since

\begin{equation}
 \sum_{i=1}^{65} \sum_z (w_A(F_i)P_A(z)+w_U(F_i)P_U(z)) = 65
\end{equation}

\begin{figure*}
\includegraphics[width=160mm,keepaspectratio]{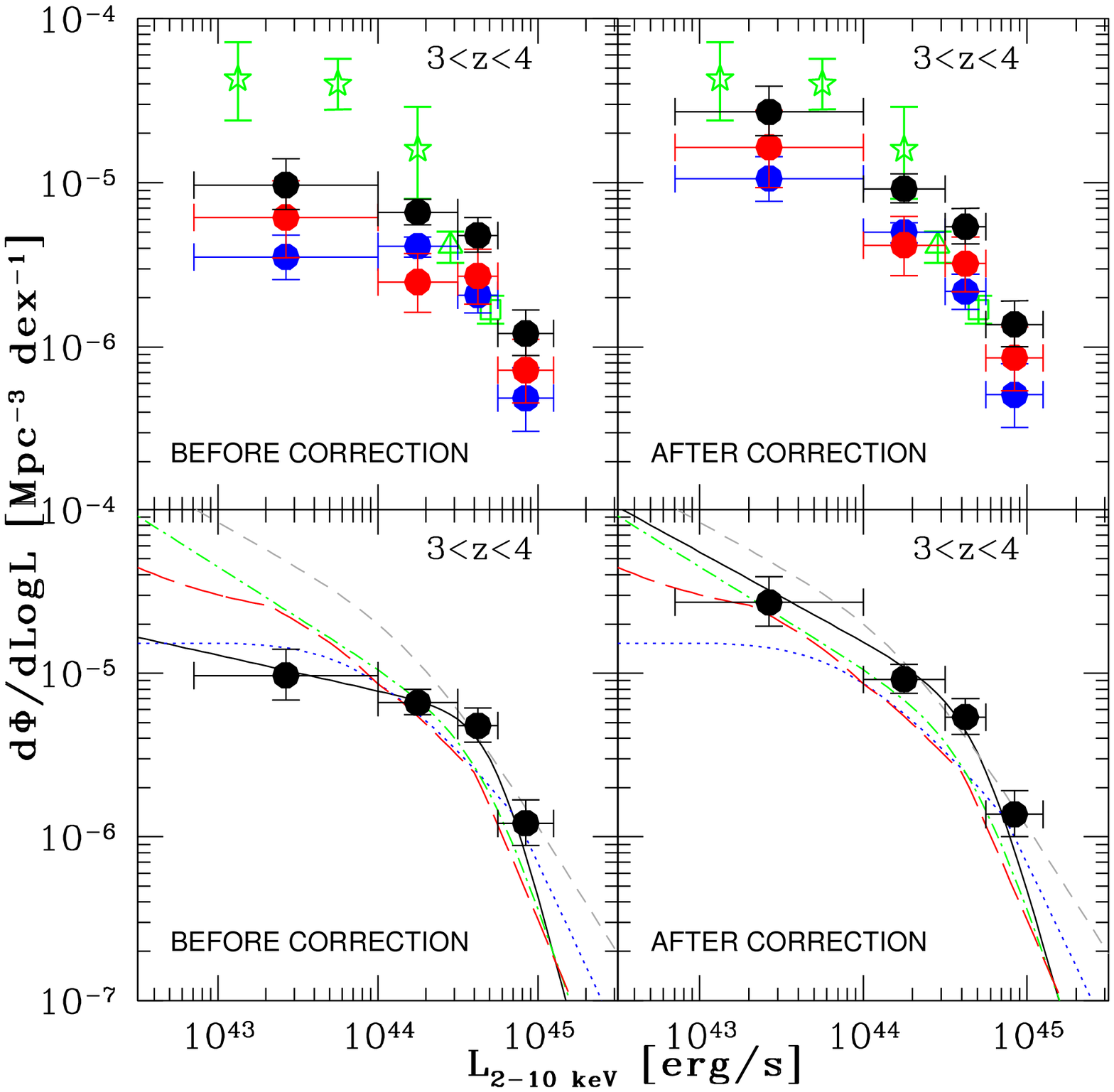}
\caption{ HXLF in a representative redshift bin ($3<z<4$) computed before (\textit{left panels}) and after (\textit{right panels}) the correction for incompleteness. In the upper panels, the HXLF of U (blue circles) and A (red circles) subsamples are plotted to show the differential correction. The total HXLF (black circles) is computed following Eq. \ref{HXLFsum}. Results from \citet[XMM-COSMOS, green square]{Brusa09} \citet[\textit{Chandra}-COSMOS, green triangle]{Civano11} and \citet[4 Ms CDF-S, green stars]{Fiore12} in the 4 Ms CDF-S, are also shown for comparison. In the lower panels, we compare the HXLF with results from literature. The LDDE model by \citet{Hasinger05}, where obscured (Compton-thin) AGN and an exponential decline in the space density are accounted following \citet{Gilli07}, is shifted to hard band and plotted as a blue dotted line. We also show models by \citet[LDDE, grey short-dashed line]{LaFranca05}, \citet[LADE, green dot-dashed line]{Aird10} and \citet[LDDE, red long-dashed line]
{Ueda14} and added our best-fitting model (PDE, black solid lines) before and 
after the correction.} \label{fig6}
 \end{figure*}

The effect of this correction is shown in Fig.~\ref{fig6}, where the HXLF is also plotted separately for the two absorption-based subsamples (upper panels). As expected, incompleteness is more severe at low X-ray luminosity, usually associated to optical faintness. Results from \citet[XMM-COSMOS, green square]{Brusa09} \citet[\textit{Chandra}-COSMOS, green triangle]{Civano11} and \citet[4 Ms CDF-S, green stars]{Fiore12} in the 4 Ms CDF-S, are also shown for comparison. In particular, \cite{Fiore12} selected a sample of high redshift objects using spectroscopic and photometric redshifts, and a colour selection. \cite{Vito13} applied a different selection, based on X-ray detection only, and a careful verification of the redshifts on the same field, which resulted in the discrepancy at the faint end of the luminosity function, dominated by the CDFS sample, when no completeness correction is applied. After the inclusion of completeness, our estimate gets closer to the \cite{Fiore12}.
Models from \citet[LDDE]{Hasinger05}, \citet[LDDE]{LaFranca05}, \citet[LADE]{Aird10} and \citet[LDDE]{Ueda14} are also plotted. Before the completeness correction, the faint end of the HXLF shows a flattening similar to that of the \citet{Hasinger05} model, but the corrected HXLF has a much steeper slope.

\begin{figure*}
\includegraphics[width=160mm,keepaspectratio]{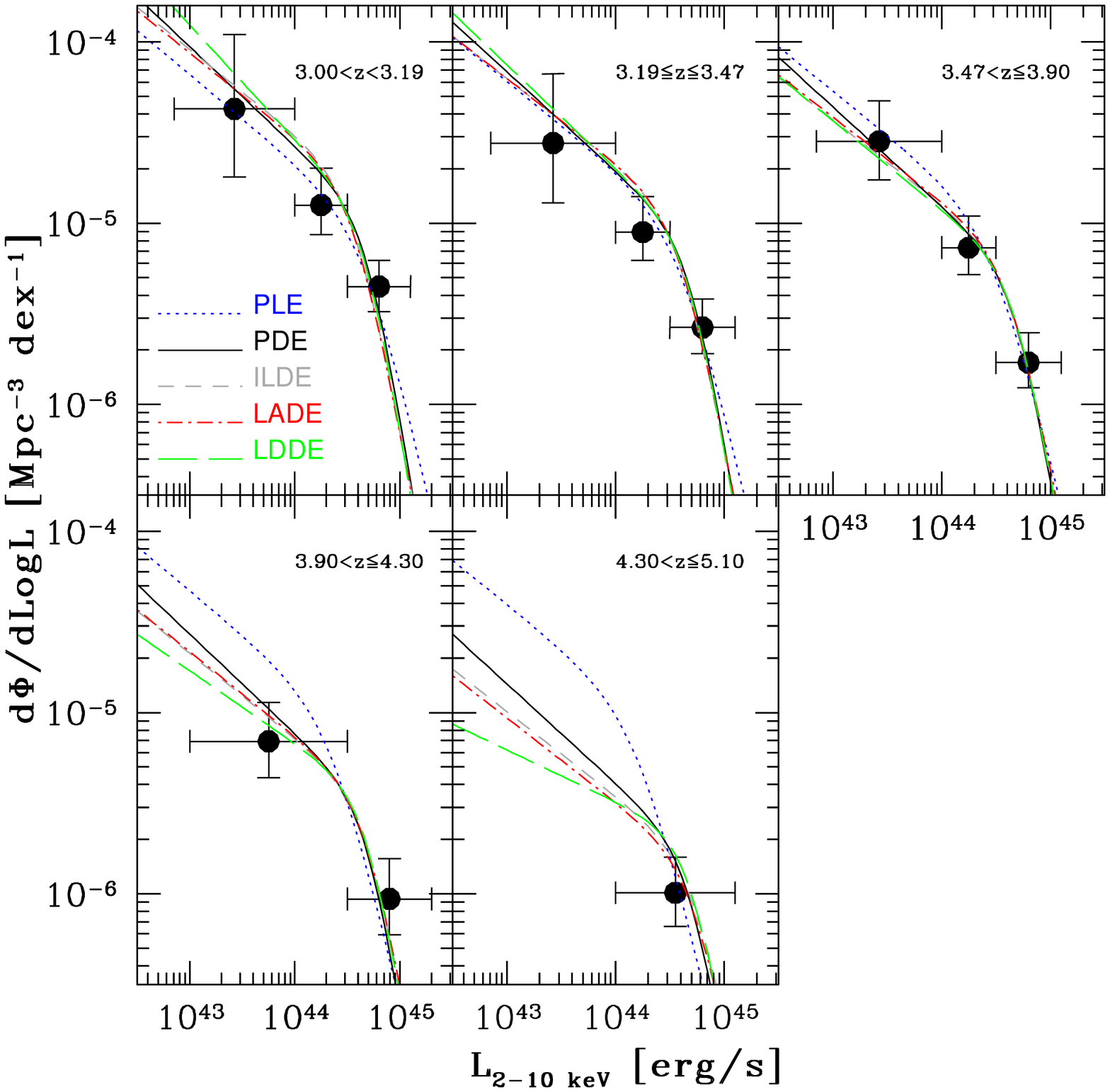}
\caption{ Same as Fig.~\ref{fig5} but including the correction for redshift incompleteness (see \S~\ref{complete}). } 
\label{fig7}
 \end{figure*}

We computed the binned HXLF and repeated the fitting procedure described in \S~\ref{maxlik} including the effect of incompleteness (Fig.~\ref{fig7}). Best-fitting parameters are reported in Tab.~\ref{tab3}.

\begin{table*}
\caption{Best-fit parameters to the HXLF, including the correction for redshift incompleteness.}\label{tab3}
\begin{tabular}{|r|r|r|r|r|r|r|r|r|}
\hline
  \multicolumn{1}{|r|}{ MODEL } &
  \multicolumn{1}{|r|}{{\bf $A $ }} &
  \multicolumn{1}{r|}{{\bf $L_*$}} &
  \multicolumn{1}{r|}{{\bf $\gamma_1$}} &
  \multicolumn{1}{r|}{{\bf $\gamma_2$}} &
  \multicolumn{1}{r|}{{\bf $p_{lum}$}} &
  \multicolumn{1}{r|}{{\bf $p_{den}$}} &
   \multicolumn{1}{r|}{{\bf $\beta$}} &
    \multicolumn{1}{r|}{ 2DKS}        \\
      (1)   &  (2)   & (3)  & (4)  &  (5)  &  (6)    & (7)    & (8) & (9)  \\

  \hline
  PLE   &  $1.00^{+0.10}_{-0.10}$ & $5.22_{-2.43}^{+2.91}$ &  $0.49_{-0.28}^{+0.19}$  &  $2.55_{-0.55}^{+0.79}$  &  $-3.19_{-0.67}^{+0.61}$  & ---    & ---  & $0.20$   \\
  PDE   &  $1.25^{+0.13}_{-0.13}$ & $5.03_{-1.55}^{+1.37}$ &  $0.55_{-0.19}^{+0.14}$  &  $3.62_{-0.86}^{+1.13}$  &  ---    & $-5.68_{-0.90}^{+0.87}$  & ---  & $0.44$   \\
  ILDE  &  $2.00^{+0.21}_{-0.21}$  & $3.63_{-1.61}^{+2.11}$  &  $0.47_{-0.22}^{+0.25}$  &  $3.15_{-0.70}^{+1.22}$  &  $1.19_{-1.29}^{+1.52}$   & $-7.19_{-2.06}^{+1.88} $ & ---  & $0.20$   \\
    LADE  &  $1.84^{+0.19}_{-0.19}$  & $3.68_{-1.61}^{+2.08}$  &  $0.47_{-0.25}^{+0.19}$  &  $3.17_{-0.71}^{+1.22}$  &  $1.12_{-1.25}^{+1.45}$   & $-0.65_{-0.18}^{+0.17} $ & ---  & $0.23$ \\
  LDDE  &  $1.19^{+0.11}_{-0.11}$ & $4.92_{-1.53}^{+1.37}$ &  $0.66_{-0.21}^{+0.17}$  &  $3.71_{-0.84}^{+1.12}$  &  ---    & $-6.65_{-1.32}^{+1.28}$ & $2.40_{-2.31}^{+2.33}$  & $0.27$   \\

  \hline

\end{tabular}

(1) model; (2) normalization in units of $10^{-5} \rmn{Mpc^{-3}}$; (3) knee luminosity in units of $10^{44}\rmn{erg}\,\rmn{s^{-1}}$; (4) faint and (5) bright end slope; (6) luminosity and (7) density evolutionary factor; (8) luminosity-dependency factor of the density evolution; (9) two dimensional Kolmogorov-Smirnov test probability.
\end{table*}

\section{Obscured AGN fraction}\label{obsfrac}

\begin{figure}
\includegraphics[width=80mm,keepaspectratio]{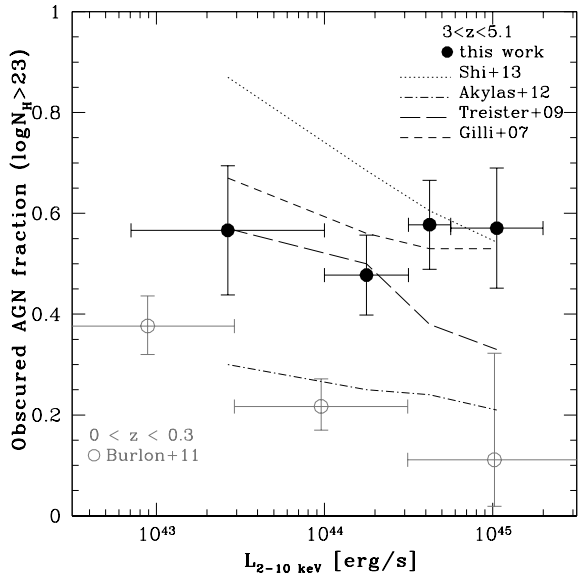}

\caption{Obscured AGN fraction as a function of luminosity. The dotted, dot-dashed, long-dashed and short-dashed lines are the predictions from the X-ray background synthesis models by \citet{Shi13}, \citet{Akylas12}, \citet{Treister09} and \citet{Gilli07}, respectively, computed at the median of the considered redshift range.}
\label{F23}
 \end{figure}

Having computed in \S~\ref{HXLF} the binned HXLF separately for the U ($\phi^U$) and A ($\phi^A$) subsamples, we can estimate the obscured AGN fraction ($F_{23}$) in a $\Delta \rmn{logL_X}-\Delta z$ bin as

\begin{equation}\label{F23eq}
 F_{23} = \frac{\phi^A}{\phi^A+\phi^U}
\end{equation}

where the subscript is added to stress that the separation between the two subsample is at $\rmn{logN_H}=23$. However, by using the same $\Delta \rmn{logL_X}-\Delta z$ bins as in \S~\ref{HXLF}, the errors on $F_{23}$ resulted too large to draw reasonable conclusions, because of the low number of sources (especially belonging to the A subsample) in such narrow bins. We chose to compute $F_{23}$ over the entire redshift range $3<z<5.1$ in four luminosity bins (Fig.~\ref{F23}).

Since the \cite{Page00} method assumes that the luminosity function does not vary in the considered bins, as discussed in \S~\ref{method}, the result we present in this section should be considered as a rough estimate of the average obscured AGN fraction in that redshift range. In \S~\ref{spden} we will derive the space density of high-redshift AGN using the $1/V_{max}$ method. The proper way to estimate $F_{23}$ would be to use the space density itself instead of $\phi$ in Eq.~\ref{F23eq}. However, obscured sources are 
on average 
closer to the flux limit of a survey than unobscured AGN at a given luminosity. The $1/V_{max}$-related bias reported in \S~\ref{method} is therefore more effective for the A subsample than for the U subsample, and this fact would strongly affect the evaluation of the obscured AGN fraction, especially at low luminosities. All values of $F_{23}$ were obtained after the correction for redshift incompleteness. We checked that no significant difference would be obtained if we had not applied that correction.

\section{Space density}\label{spden}
\begin{figure}
\includegraphics[width=80mm,keepaspectratio]{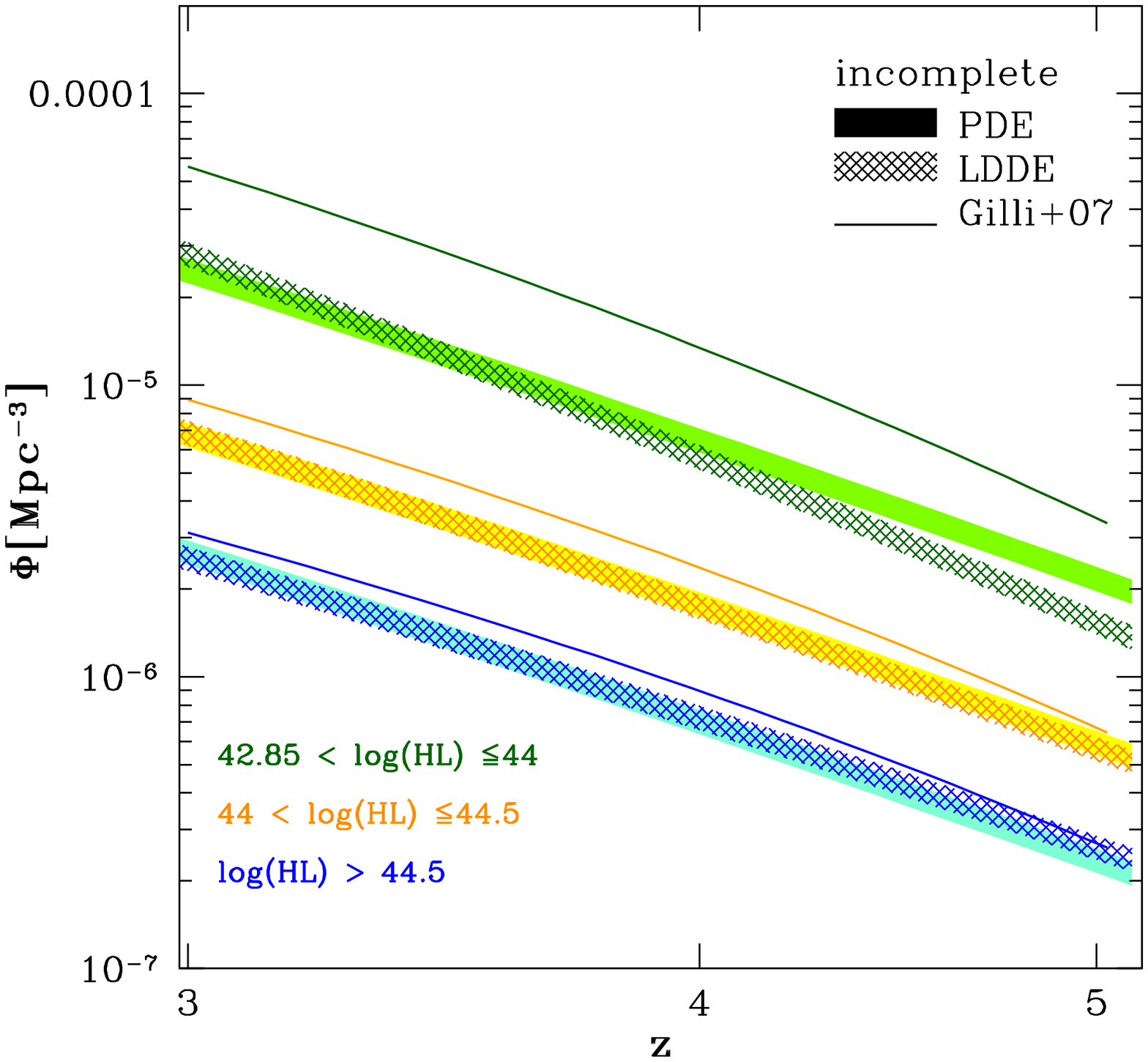}
\includegraphics[width=80mm,keepaspectratio]{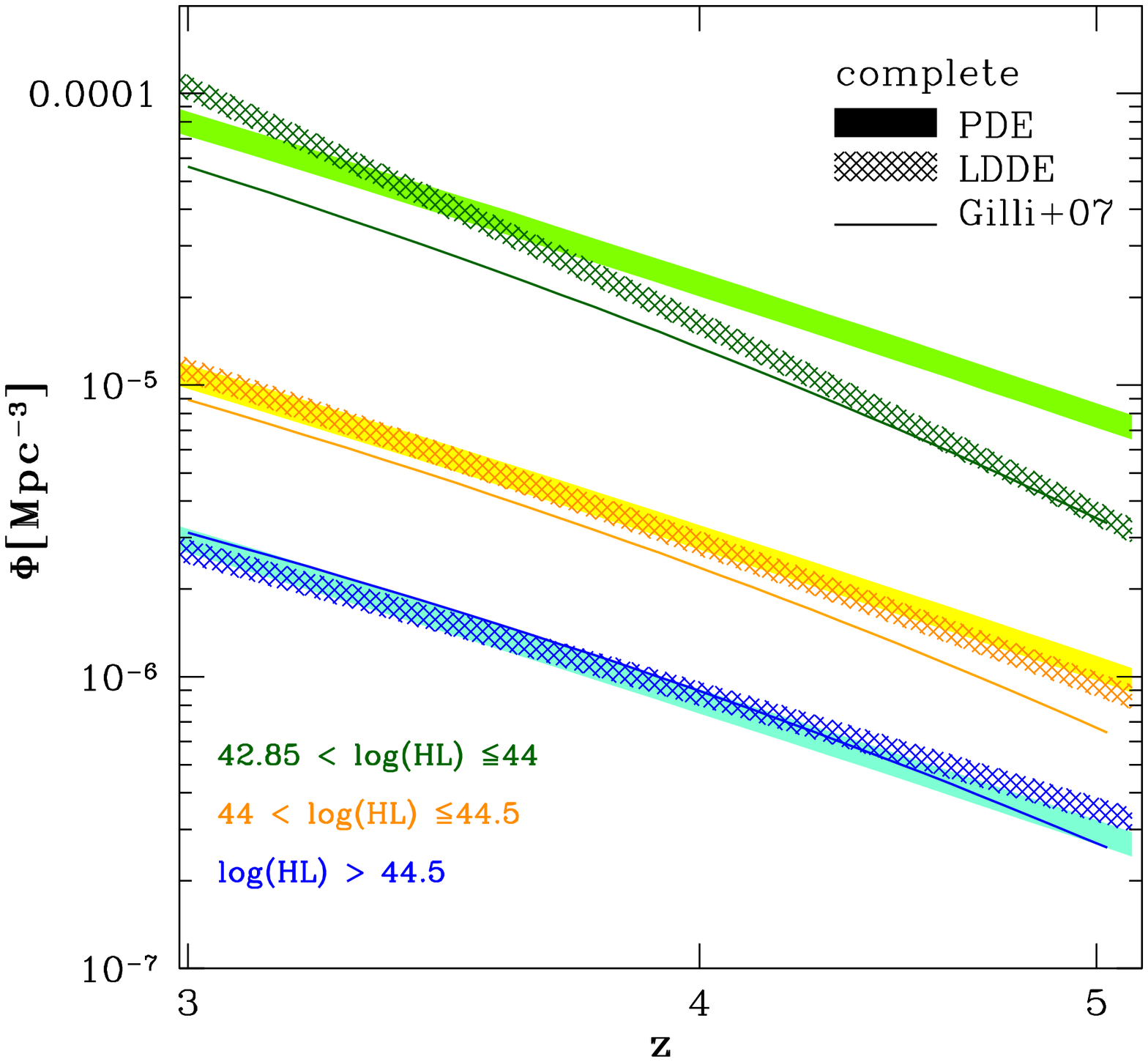}

\caption{ Space density in three different, colour-coded luminosity bins, estimated from the best-fitting PDE (filled, lighter stripes) and LDDE (gridded, darker stripes) models. In the upper panel, we assumed $\Theta=1$ (i.e. no redshift completeness correction), while in the lower panel we applied the correction for redshift incompleteness (as in \S~\ref{complete}). 
The three solid lines are the predictions from the \citet{Gilli07} X-ray background synthesis model computed including also the most Compton-thick AGN ($N_H>10^{25}\rmn{cm^{-2}}$).  }
\label{fig8a}
 \end{figure}

 The space density expected from the best-fitting PDE and LDDE models is plotted as a function of redshift in Fig.~\ref{fig8a} in three different luminosity bins, before (upper panel) and after (lower panel) correcting for the redshift incompleteness. The width of the stripes accounts only for the uncertainties on the normalization of each model (i.e. on the number of sources). 
 We also show the expectations from the \cite{Gilli07} X-ray background synthesis model, including also the most Compton-thick AGN ($N_H>10^{25}\rmn{cm^{-2}}$) and adopting a high-redshift decline of the AGN space density\footnote{http://www.bo.astro.it/$\sim$gilli/xvol.html}. 
 
A binned representation of the space density $\Phi$ of an extragalactic population of objects can be derived using the $1/V_{max}$ method (\citealt{Schmidt68,Avni80}):
 
 \begin{equation}
  \Phi=\frac{\rmn{d}N}{\rmn{d}V}=\sum_{i=1}^{N}\frac{1}{V_{max,i}}=\sum_{i=1}^{N}\frac{1}{\int \Omega\Theta\frac{\rmn{d}V}{\rmn{d}z}\,\rmn{d}z}
 \end{equation}
where $N$ is the number of objects in the $\Delta\rmn{Log}L - \Delta z$ bin of interest and $\Theta=\Theta(z,L_{x,i},N_H)$ is the completeness factor, as previously defined.

We computed the binned space density of $\rmn{logL_X}>44.15$ AGN as $\Phi=\Phi_{U}+\Phi_{A}$, where $\Omega^{\rmn{U}}$ and $\Omega^{\rmn{A}}$ are used to compute $\Phi_{U}$ and $\Phi_{A}$, respectively. 
Similarly to \cite{Hiroi12}, errors are estimated as

\begin{equation}
 \delta\Phi_U=\sqrt{\sum_{i=1}^{N_U}\frac{1}{V_{max,i}^2}}
\end{equation}
where $N_U$ is the number of sources in the U subsample, for $\Phi_U$ and likewise for $\Phi_A$. Then, they are propagated to obtain the error on $\Phi$. The results are plotted in Fig.~\ref{fig9} and will be discussed in \S~\ref{ev_spden}.

 In principle, we could also use the \cite{Page00} method to derive a binned HXLF in a given luminosity bin and then multiplying it by the corresponding $\Delta\rmn{LogL}$ to obtain $\Phi$. However, we computed the space density in a luminosity bin ($\rmn{logL_X}>44.15$) even larger that those used in \S~\ref{HXLF} and the assumption that $\phi$ is not varying in that bin would strongly affect the evaluation of the binned points. Instead, since the fluxes corresponding to such luminosities are expected to be higher than the flux limit of the surveys, we can safely use the $\frac{1}{V_{max,i}}$ method (see \S~\ref{method}).

\section{Discussion}

In this section we discuss the results derived for the HXLF, AGN absorbed fraction and space density. We will also compare our findings with those from literature.

\subsection{Evolution of the HXLF}

Before the correction for redshift incompleteness, the 2DKS test returns similar high values for the PDE, ILDE, LADE and LDDE models (Tab.~\ref{tab2}). However, the common parameters are very similar and, since the ILDE, LADE and LDDE models have an additional free parameter consistent with zero (see Tab.~\ref{tab2}), we conclude that these three models mimic the behaviour of the PDE model. Indeed, the PDE, ILDE and LADE models plotted in Fig.~\ref{fig5} are almost completely overlapping. As for the LDDE model, we note that the evolution dependency on the luminosity is more effective at low luminosities in changing the shape of the HXLF, because of the rather flat faint-end slope. Therefore, we ascribe the weak constraints we found on the $\beta$ parameter to the poor sampling of the low-luminosity regime. 
 
After the correction for redshift incompleteness, the PDE model has the highest 2DKS test value. The LDDE model is still consistent with the data. 
The best-fitting ILDE and LADE models are again consistent with no luminosity evolution, although less significantly than in \S~\ref{maxlik} (see Tab.~\ref{tab3}). According to the 2DKS probability, these models, as well as the LDDE and PLE ones, are acceptable with a much smaller significance than in the previous case..
 
We conclude that the evolution of the $z > 3$ HXLF is dominated by a negative density term. A luminosity evolution term  which is present in the other parameterization of the luminosity function is either inconsistent (i.e. PLE) or not required (ILDE,LADE and LDDE) by the data.

Recently, \cite{Ueda14} presented the X-ray luminosity function of AGN from $z=0$ to 5 and chose a LDDE model to analitically describe the evolution, fixing the model parameters at high redshift ($z>3$). Instead, we focused on the particular redshift range $3<z<5.1$. Therefore the best-fitting parameters are not heavily weighted by lower redshift data, were much larger samples of AGN are available. We also have not fixed any model parameter.

\subsection{Evolution of the obscured AGN fraction}
The obscured AGN fraction at $3\leq z \lesssim 5$ seems to be constant with luminosity (a simple $\chi^2$ fit to the filled circles in Fig.~\ref{F23} returns a value of $F_{23}=0.54\pm0.05$). This is in contrast to the decreasing obscured fraction with luminosity reported in many works, mainly at lower redshift, using X-ray \citep[e.g.][but see also \citealt{Merloni14}]{Steffen03,Ueda03,LaFranca05,Treister06,Hasinger08,Ebrero09, Brusa10,Ueda14} and optical/IR data \citep[e.g.][]{Simpson05, Treister08, Lusso13}. \cite{Kalfountzou14} found a similar decreasing trend of the obscured fraction with luminosity in their sample of high-redshift X-ray selected AGN. However, the luminosities they consider are as observed (i.e. not corrected for absorption), therefore the trend could be at least partially driven by intrinsically-luminous, obscured AGN which are counted in the low observed-luminosity bins.

Heavily-obscured ($\rmn{logN_H}\gtrsim23.5$), low luminosity ($\rmn{logL_X}\lesssim44$) AGN at $z>3$ cannot be detected even by the deepest X-ray surveys. Indeed, in our sample there is a clear deficiency of such objects with respect to more luminous ($\rmn{logL_X}\gtrsim44$) AGN affected  by a similar level of obscuration in our sample, as it can be seen in Fig.~\ref{fig2} (lower panel). The obscured AGN fraction could then be larger than our estimates, especially in the first luminosity bin of Fig.~\ref{F23}. We note that if we assumed $\rmn{logN_H}=22$ to be the column density threshold dividing obscured and unobscured sources, as usually done in literature, the obscured AGN fraction would result even larger than the value we found.

We also compared our points with the predictions from the \citet[]{Gilli07}, \citet[]{Treister09}\footnote{http://agn.astroudec.cl/j\_agn/main}, \citet[]{Akylas12}\footnote{http://indra.astro.noa.gr/xrb.html}  and \citet[]{Shi13}\footnote{http://5muses.ipac.caltech.edu/5muses/\\EBL\_model/num\_den\_red.html} X-ray background synthesis models (see Fig.~\ref{F23}). The same definition of the obscured AGN fraction as in Eq.~\ref{F23eq} was adopted to compute the predictions.
At low luminosities, our points are consistent with the \cite{Treister09} and \cite{Gilli07} models, while at high luminosities we found a good agreement with the \cite{Gilli07} and \cite{Shi13} predictions. 

When comparing our points (filled circles in  Fig.~\ref{F23}) with results in the Local Universe from \citet[]{Burlon11}, who studied a complete sample of AGN detected by \textit{Swift}--BAT, we found a positive evolution of the obscured AGN fraction from $z=0$ to $z>3$, which is stronger at high luminosities  ($\rmn{logL_X}>44$), even considering that the low luminosity bin at $z >3$ could be affected by some incompleteness. This result agrees with the increasing fraction of absorbed AGN with redshift reported in literature \citep[e.g.][]{LaFranca05, Akylas06, Treister06, Hasinger08, Burlon11, Ueda14}. This finding fits well in a wider scenario \citep[e.g.][]{Hopkins08,Hickox09} in which luminous ($\rmn{logL_X}>44$) AGN are mainly triggered by wet (i.e. gas-rich) mergers \citep[e.g.][]{DiMatteo05, Menci08}. In this case the gas may accrete chaotically, producing large covering factors. The obscured accretion phase is then rapidly terminated by the powerful nuclear radiation, which sweeps away the gas, 
allowing the QSO to reveal itself as unobscured. Since the gas fraction is thought to increase 
with redshift \citep[e.g.][]{Carilli13}, a longer obscured phase and larger 
covering angles are expected at higher redshift, causing a positive evolution of the obscured AGN fraction from the Local to the high-redshift Universe. Low-luminosity ($\rmn{logL_X}<44$) AGN may instead be triggered preferentially by smooth, secular processes \citep[e.g.][]{Elbaz11} where gas accretion is more symmetric, and the classical ``unification scheme'' \citep[e.g.][and reference therein]{Urry95}, in which obscuration is only due to the geometry of the system with respect to the line of sight, holds. In this case, no strong evolution of the fraction of obscured AGN is expected with redshift.

\begin{figure}
\includegraphics[width=80mm,keepaspectratio]{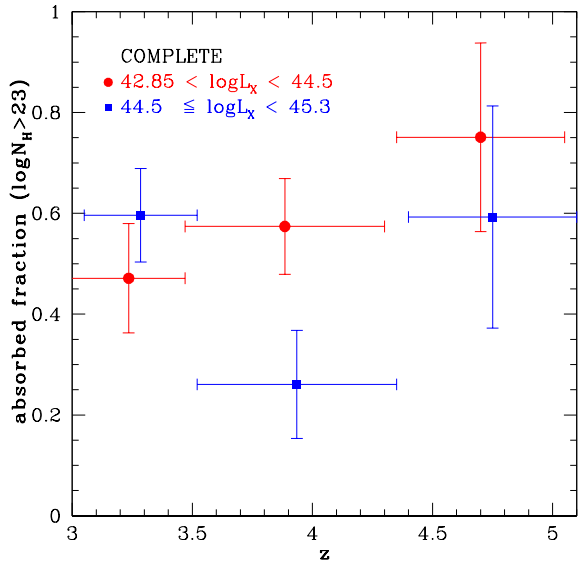}

\caption{Obscured AGN fraction as a function of redshift for two luminsity ranges. }
\label{F23_L}
 \end{figure}

In order to search for a possible evolution from $z=3$ to $z=5$, in Fig.~\ref{F23_L} we plotted the obscured AGN fraction against redshift for two luminosity bins. The large error bars, the narrow redshift range, the relatively low number of (obscured) sources and the caveats reported in \S~\ref{obsfrac} prevent us from drawing strong conclusions. We note that an absence of evolution at high redshift would be in agreement with \cite{Hasinger08}, who suggested that the Type-2 AGN fraction saturates at $z\ge2$.

\subsection{Evolution of the space density}\label{ev_spden}
  Recently \cite{Ueda14} argued that, at $z > 3$, X-ray selected AGN experience an ``up-sizing" evolution (i.e. the space density of low luminosity AGN declines with increasing redshift at a slower rate than for high luminosity  objects).
 There is no evidence of such an effect in our sample, if anything there may be hints of an opposite trend (see Fig.~\ref{fig8a}, dashed lines).
  The origin of the discrepancy could be due to the different sample selection at high-z and/or to the details of the choice of the parameters which are fixed 
  in the \cite{Ueda14} fits.

\begin{figure}
\includegraphics[width=80mm,keepaspectratio]{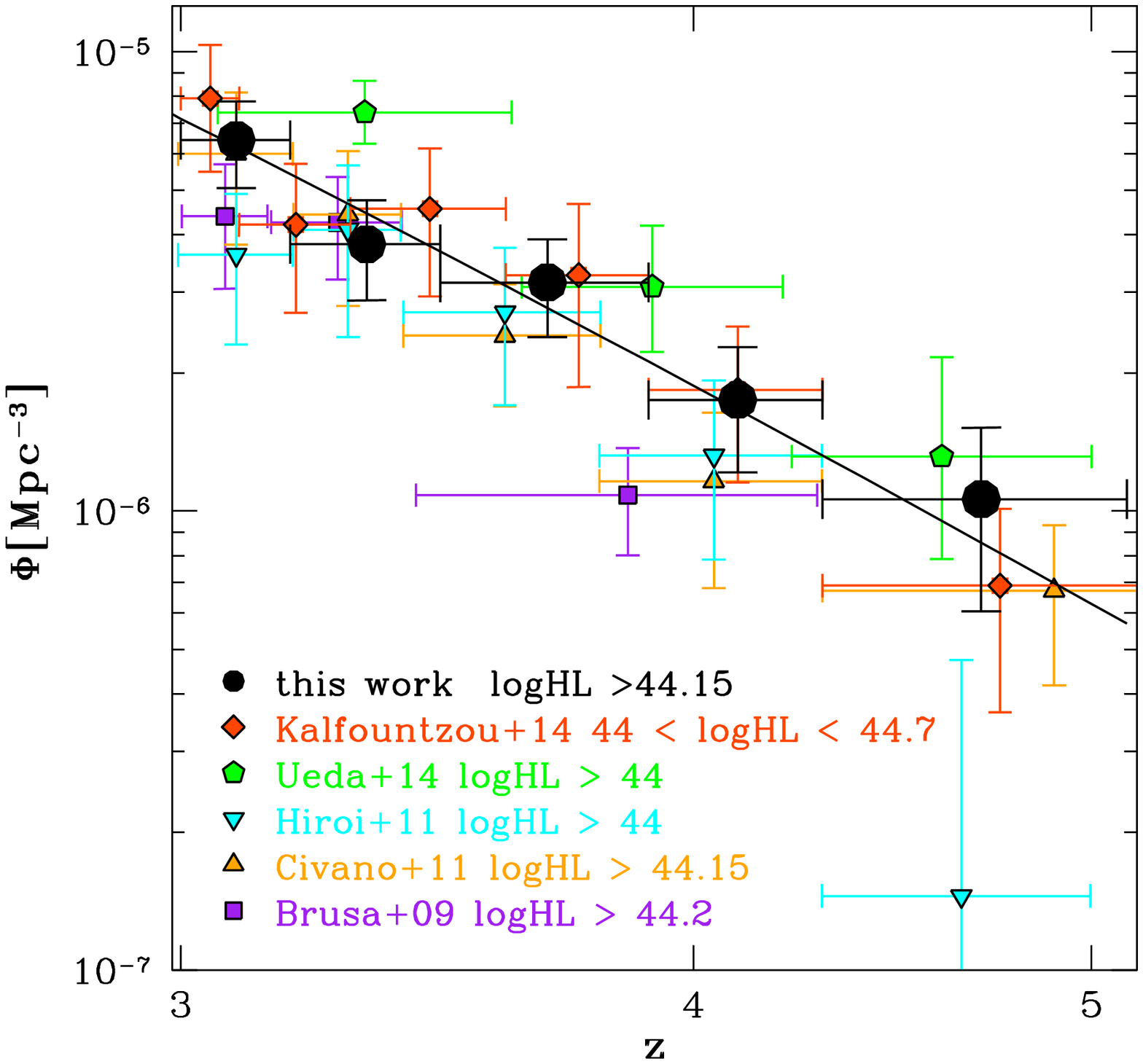}
\includegraphics[width=80mm,keepaspectratio]{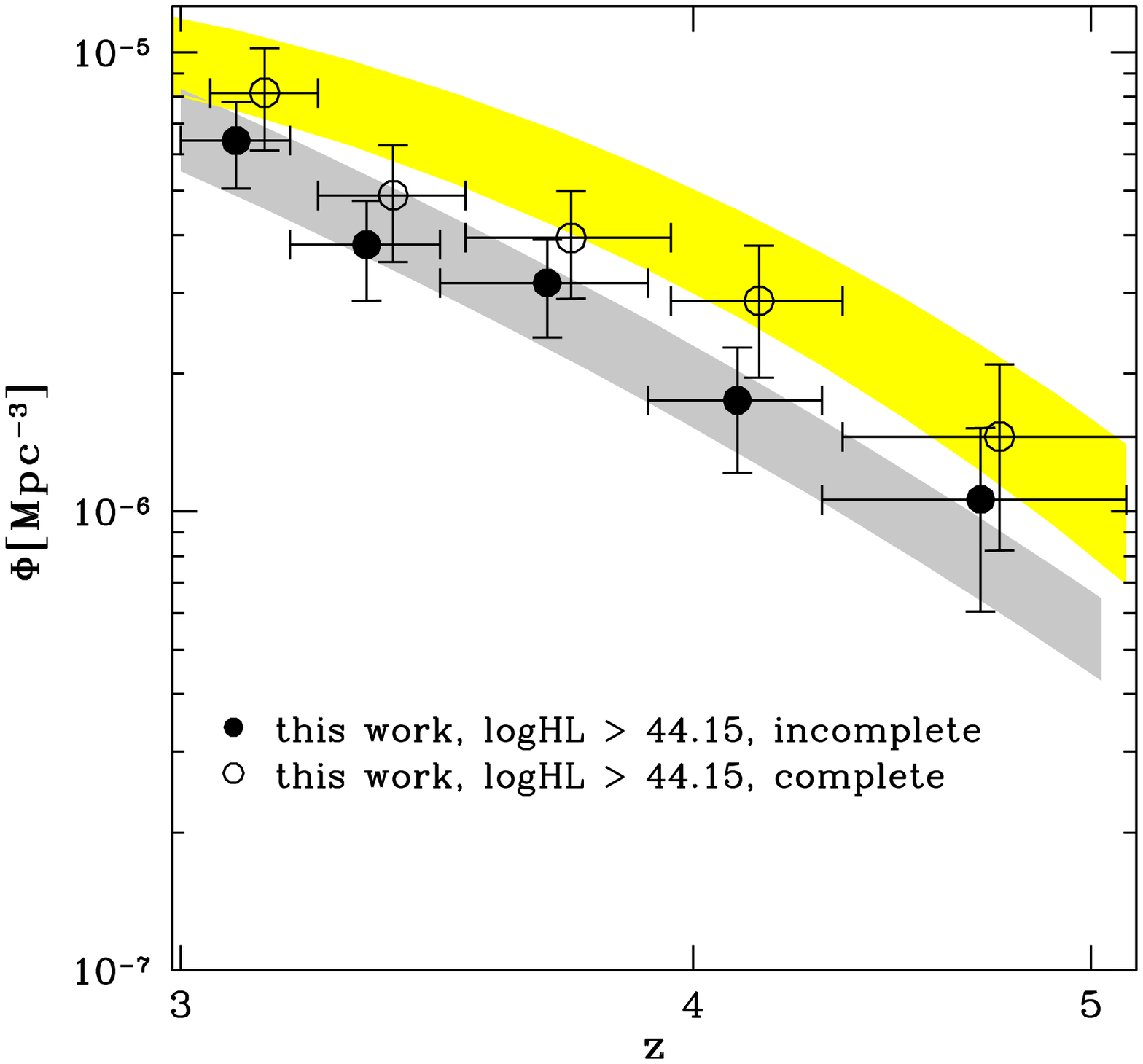}

\caption{ Upper panel: our estimate of the AGN space density ($\rmn{LogL}>44.15$, not corrected for incompleteness; black circles) as a function of redshift is compared with results from \citet[$44<\rmn{LogL}<44.7$, red diamonds]{Kalfountzou14}, \citet[$\rmn{LogL}>44$, green pentagons]{Ueda14}, \citet[$\rmn{LogL}>44$, cyan downward-pointing triangles]{Hiroi12}, \citet[$\rmn{LogL}>44.15$, orange upward-pointing triangles]{Civano11} and \citet[$\rmn{LogL}>44.2$, purple squares]{Brusa09}. The solid line is the best-fitting PDE model. Lower panel: AGN space density before (filled circles) and after (empty circles) the correction for redshift incompleteness.
The grey and yellow stripes are the expectations from the X-ray background synthesis models by \citet{Gilli07} and \citet{Shi13}, respectively, computed at $\rmn{LogL}>44.15$ between the two extreme cases of accounting only for Compton-thin ($N_H<10^{24}\,\rmn{cm^{-2}}$; lower bounds) and including also Compton-thick (upper bounds) AGN. }
\label{fig9}
 \end{figure}

 In Fig.~\ref{fig9} we compare our results at $\rmn{LogL_X}>44.15$, not corrected for redshift incompleteness, with those from \citet[$44<\rmn{LogL_X}<44.7$]{Kalfountzou14}, \citet[$\rmn{LogL_X}>44$]{Ueda14}, \citet[$\rmn{LogL_X}>44$]{Hiroi12}, \citet[$\rmn{LogL_X}>44.15$]{Civano11} and \citet[$\rmn{LogL_X}>44.2$]{Brusa09}, and with the predictions from the X-ray background synthesis models by \citet{Gilli07} and \citet{Shi13}. In particular, \cite{Hiroi12} jointly fitted their and the \cite{Civano11} points with a powerlaw in the form $\Phi\propto(1+z)^{p}$, finding a decline in the space density of high-redshift AGN with a slope of $p=-6.2\pm0.9$. Our best-fitting PDE model returns a very similar evolution (see Tab.~\ref{tab2} and \ref{tab3}), although a larger dataset (which includes the \citeauthor{Hiroi12} sample) was used and a different method (i.e. maximum likelihood fit to unbinned data in the $z-\rmn{logL_X}$ space) was adopted.

\section{Conclusions}
We presented a sample of 141 soft-band X-ray selected AGN at $3<z\lesssim 5$ to date, using data from the 4 Ms CDF-S, \textit{Chandra}-COSMOS, XMM-COSMOS and SXDS surveys, characterised by a redshift completeness of $>90$ per cent. More than a half ($\sim 55$ per cent) of the objects in the sample has a spectroscopic redshift; a photometric redshift was adopted for the remaining $\sim 45$ per cent of the sources. We defined two absorption-based subsamples, assuming a dividing value of $\rmn{logN_H}=23$. In order to constrain the evolution of the HXLF at high redshift, we fitted different analytical models to the unbinned $z-\rmn{logL_X}$ pairs through a maximum likelihood procedure, taking into account the effect of the different spectral shape between unobscured and obscured AGN on the sky coverages of the surveys.The main results of this work are the following:

 \begin{enumerate}
  \item The evolution of the HXLF at $3<z\lesssim5$ is dominated by a negative density evolution term. A Pure Density Evolution (PDE) model best represents our data. A Luminosity And Density Evolution (LADE) and an Independent Luminosity and Density Evolution (ILDE) models mimic the behaviour of the PDE model in this redshift range. 

  \item We estimated the fraction of $z>3$ AGN obscured by a column density $\rmn{logN_H}>23$ to be $0.54\pm 0.05$. In contrast with many other works performed on lower-redshift data, there is no strong evidence for an anti-correlation between the obscured AGN fraction and luminosity in the redshift range probed by this work. However, the most obscured, low-luminosity AGN are probably not detected; this bias could hide a possible anti-correlation. Comparing our points with previous finding in the Local Universe, we confirm the positive evolution of the obscured AGN fraction with redshift reported by many authors. 
  
  \item The space density of luminous AGN decreases by a factor of $\sim 10$ from $z=3$ to 5. By using a larger sample and a different procedure (maximum likelihood fit to unbinned data), we then confirm the decline found by other authors in similar ranges of redshift and luminosity. No hint of an ``up-sizing" was found. However, larger samples of AGN, especially of low luminosity ($\mathrm{logL_X}<44$) and at $z>4$, are necessary to constrain the space density of high-redshift AGN and will be provided by future surveys like the additional \textit{Chandra} 3 Ms observation in the CDFS (PI: N. Brandt) and the \textit{Chandra}-COSMOS Legacy survey (PI: F. Civano).
 \end{enumerate}

\section*{Acknowledgments}
We acknowledge financial support from INAF under the contracts PRIN-INAF-2011 and PRIN-INAF-2012. MB acknowledges support from the FP7 Career Integration Grant “eEASy” (“SMBH evolution through cosmic time: from current surveys to eROSITA-Euclid AGN Synergies, CIG 321913). We thank the anonymous referee for the useful comments and suggestions.  We are grateful to Y. Ueda, T. Miyaji, G. Lanzuisi, D. Burlon and E. Kalfountzou for providing their results.

\bsp

\label{lastpage}

\end{document}